\documentclass[draftclsnofoot, onecolumn, letterpaper, romanappendices]{IEEEtran}

\usepackage{ifpdf}

\usepackage[noadjust]{cite}
\ifCLASSINFOpdf
   \usepackage[pdftex]{graphicx}
\else
\fi
\usepackage[cmex10]{amsmath}

\usepackage{algorithmic}

\usepackage{array}
\usepackage{mdwmath}
\usepackage{mdwtab}
\usepackage{eqparbox}
\usepackage[font=footnotesize]{subfig}
\usepackage{hyperref}

\newtheorem{theorem}{Theorem}

\newtheorem{lemma}[theorem]{Lemma}
\newtheorem{corollary}[theorem]{Corollary}

\newtheorem{define}{Definition}
\newtheorem{proposition}{Proposition}
\newtheorem{example}{Example}
\newtheorem{remark}{Remark}

\newcommand{\field}[1]{\mathbb{#1}}
\newcommand{\N}{\field{N}} 
\newcommand{\R}{\field{R}} 
\renewcommand{\Re}{\R} 
\newcommand{\E}{\mathsf{E}} 

\newcommand{\Iscr}{{\cal I}}

\newcommand{\Nscr}{{\cal N}}

\newcommand{\tsnr}{{\text{snr}}}
\newcommand{\tvar}{{\text{Var}}}

\newcommand{\bea}{\begin{eqnarray}}
\newcommand{\eea}{\end{eqnarray}}
\newcommand{\beas}{\begin{eqnarray*}}
\newcommand{\eeas}{\end{eqnarray*}}
\usepackage{amsfonts}

\begin{document}
\title{Pointwise Relations between Information and Estimation in Gaussian Noise}
%

\author{Kartik Venkat\IEEEauthorrefmark{1}, Tsachy Weissman\IEEEauthorrefmark{1}
\thanks{\IEEEauthorrefmark{1} Stanford University. Email: kvenkat@stanford.edu, tsachy@stanford.edu} 
}


%
%


\maketitle

\begin{abstract}
Many of the classical and recent relations between information and estimation in the presence of Gaussian noise can be viewed as identities between expectations of random quantities. These include  the I-MMSE relationship of Guo et al.; the relative entropy and mismatched estimation relationship of Verd\'{u}; the relationship between causal estimation and mutual information of Duncan, and its extension to the presence of feedback by Kadota et al.; the relationship between causal and non-casual estimation of Guo et al., and its mismatched version of Weissman. We dispense with the expectations and explore the nature of the pointwise relations between the respective random quantities. The pointwise relations that we find are as succinctly stated as - and give considerable insight into -  the original expectation identities. 

As an illustration of our results, consider Duncan's 1970 discovery that the mutual information is equal to the causal MMSE in the AWGN channel, which can equivalently be expressed saying that the difference between the input-output information density and half the causal estimation error is a zero mean random variable (regardless of the distribution of the channel input). We characterize this random variable explicitly, rather than merely its expectation. Classical estimation and information theoretic quantities emerge with new and surprising roles. For example, the variance of this random variable turns out to be given by the causal MMSE (which, in turn, is equal to the mutual information by Duncan's result).


\end{abstract}


%
\IEEEpeerreviewmaketitle

\section{Introduction} \label{sec: introduction}
The literature abounds with results that relate classical quantities in information and estimation theory. Of particular elegance are relations that have been established in the presence of additive Gaussian noise. In this work, we refine and deepen our understanding of these relations by exploring their `pointwise' properties. 

Duncan, in \cite{duncan}, showed that for the continuous-time additive white Gaussian noise channel, the minimum mean squared filtering(causal estimation) error is twice the input-output mutual information for any underlying signal distribution. Another discovery was made by Guo et al. in \cite{gsv2005}, where the derivative of the mutual information was found to equal half the minimum mean squared error in non-causal estimation. By combining these two intriguing results, the authors of \cite{gsv2005} also establish the remarkable equality of the causal mean squared error (at some `signal to noise' level snr) and the non-causal error averaged over `signal to noise' ratio uniformly distributed between 0 and snr. There have been extensions of these results to the presence of mismatch. In this case, the relative entropy and the difference of the mismatched and matched mean squared errors are bridged together: Mismatched estimation in the scalar Gaussian channel was considered by Verd\'{u} in \cite{ver_mis10}. In \cite{weissman10}, a generalization of Duncan's result to incorporate mismatch for the full generality of continuous time processes is provided. In \cite{kadota71}, Kadota et al. generalize Duncan's theorem to the presence of feedback. These, and similar interconnections between information and estimation are quite intriguing, and merit further study of their inner workings, which is the goal of this paper .

The basic information-estimation identities, such as the ones mentioned above, can be formulated as expectation identities. We explicitly characterize the random quantities involved in a pointwise sense, and in the process elicit new connections between information and estimation for the Gaussian channel. Girsanov theory and It\^{o} calculus provide us with tools to understand the pointwise behavior of these random quantities, and to explore their properties.  

The paper is organized as follows. In Section \ref{sec: main results}, we present and discuss our main results. In Section \ref{sec: revisiting the scalar setting},  we further develop and expand our results and observations for the setting of scalar random variables. The detailed proofs are provided in Section \ref{sec: proofs}. We conclude in Section \ref{sec: conclusion} with a summary of our main findings.

\section{Main Results} \label{sec: main results}

\subsection{Scalar Estimation} \label{sec: scalar estimation}

\subsubsection{Matched Case} \label{sec: scalar matched setting}
We begin by describing the problem setting. We are looking at mean square
estimation in the presence of additive white Gaussian noise. This problem is
characterized by an underlying clean signal $X$ (which follows a law $P_X$) and
its AWGN corrupted version $Y_\gamma$ measured at a given `signal-to-noise
ratio' $\gamma$, which is to say
\bea
Y_\gamma | X \sim \Nscr(\sqrt{\gamma}X,1), \label{eq: gaussian channel condition}
\eea
or equivalently
\bea
Y_\gamma | X \sim \Nscr(\gamma X, \gamma),
\eea
where $\Nscr(\mu, \sigma^2)$ denotes the Gaussian distribution with mean $\mu$
and variance $\sigma^2$.

In a communication setting, we are interested in the mutual information between the input $X$ and the output $Y_\gamma$. It quantifies the ability of the channel to convey useful information. In an estimation setting, one would be interested in using the observed output to estimate the underlying input signal while minimizing a given loss function. 
Define $\text{mmse}(\gamma)$ to be the minimum mean square error at `signal-to-noise ratio' $\gamma$
\bea
\text{mmse}(\gamma) = \E\big[(X - \E[X|Y_{\gamma}])^{2}\big]. \label{eq: mmse scalar}
\eea
Intriguing ties have been discovered between the input-output mutual information and the mean squared estimation loss for the Gaussian channel. In \cite{gsv2005}, Guo et. al. discovered the I-MMSE relationship; which tells us that for the additive Gaussian channel, the following relationship holds between the minimum mean square error and the mutual information between input $X$ and output $Y_{\tsnr}$ (the subscript making the `signal-to-noise ratio' explicit)
\bea
\frac{d}{d \tsnr} I(X;Y_\tsnr) = \frac{1}{2} \text{mmse}(\tsnr). \label{eq: I-MMSE differential form}
\eea
Writing (\ref{eq: I-MMSE differential form}) in its integral form,
\bea
I(X;Y_{\tsnr}) = \frac{1}{2}\int_0^{\tsnr} \text{mmse}(\gamma)\,d\gamma. \label{eq: i-mmse form 1}
\eea
Recall that we can express the mutual information between two random variables as
\bea
I(X;Y) = \E\bigg[\log\frac{dP_{Y|X}}{dP_{Y}} \bigg], \label{eq: I expectation form}
\eea 
where the quantity in the brackets denotes the $\log$ Radon-Nikodym derivative of the measure induced by the conditional law of $Y|X$ with respect to the measure induced by the law of $Y$. This quantity is referred to in some parts of the literature as the input-output information density $i(X,Y)$ (cf. \cite{PoPoorVerdu08}). In particular, let us look at the following additive Gaussian channel at `signal-to-noise ratio' $\gamma$,
\bea 
Y_\gamma = \gamma\,X + W_\gamma, \label{eq: Brownian motion coupling}
\eea
for $\gamma \in [0,\tsnr]$, where $W_{\cdot}$ is a standard Brownian motion
\cite{KaratzasShreve}, independent of $X$. Recall that
$W_\gamma \sim \Nscr(0,\gamma)$.

Now that $(X,Y_0^{\tsnr})$ are on the same probability space (where throughout
$Y_0^{snr}$ is shorthand for $\{ Y_\gamma , 0 \leq \gamma \leq snr \}$), it is
meaningful to interchange the expectation and integration in the right hand
side of (\ref{eq: i-mmse form 1}) to yield the following equivalent
representation of the I-MMSE result
\bea
\E\bigg[\log\frac{dP_{Y_{\tsnr}|X}}{dP_{Y_{\tsnr}}}\bigg] = \E\bigg[ \frac{1}{2}\int_0^{\tsnr} (X - \E[X|Y_{\gamma}])^{2}\,d\gamma\bigg]. \label{eq: exchange expectation and integral in I-MMSE scalar}
\eea
In other words, the I-MMSE relationship can be restated succinctly as:
\bea
\E[Z] = 0, \label{eq: zero expectation form of I-MMSE result}
\eea
where
\bea
Z = \log\frac{dP_{Y_{\tsnr}|X}}{dP_{Y_{\tsnr}}} - \frac{1}{2}\int_0^{\tsnr} (X - \E[X|Y_{\gamma}])^{2} \,d\gamma  \label{eq: Z}
\eea
denotes the ``tracking error between the information density and half the squared error integrated over snr''. But what can we say about the random variable $Z$ itself, beyond the fact that it has zero mean? Is there a crisp characterization of this random variable? The answer is captured in the following Proposition, where we present our first pointwise result. 
\begin{proposition} \label{lemma: pointwise scalar Z for Brownian motion coupling}
Assume $X$ has finite variance. $Z$, as defined in (\ref{eq: Z}), satisfies
\bea
Z = \int_0^\tsnr (X - \E[X|Y_\gamma])\cdot\,dW_\gamma \ \ \ a.s., \label{eq: pointwise scalar Z for Brownian motion coupling}
\eea 
where the integral on the right hand side of (\ref{eq: pointwise scalar Z for Brownian motion coupling}) denotes the It\^{o} integral with respect to $W_{\cdot}$.
\end{proposition}
In particular, the above characterization implies that $Z$ is a martingale, and (by virtue of having zero expectation) directly implies the I-MMSE relationship in (\ref{eq: zero expectation form of I-MMSE result}) (which is equivalent to (\ref{eq: i-mmse form 1})). Another immediate consequence of Proposition \ref{lemma: pointwise scalar Z for Brownian motion coupling} is the following:
\begin{theorem} \label{thm: variance of Z scalar for Brownian motion coupling} 
Assume $X$ has finite variance. Then
\bea
Var(Z) = \int_0^\tsnr \text{mmse}(\gamma)\,d\gamma = 2 I(X; Y_\tsnr). \label{eq: variance of Z scalar for Brownian motion coupling}
\eea  
\end{theorem}   
Thus we observe a simple characterization of the second moment of the tracking error, in terms of classical estimation and information quantities. The relationship in (\ref{eq: variance of Z scalar for Brownian motion coupling}) tells us how far apart the information density and the estimation error typically are, two quantities that we know to have equal expectations - and in particular that the variance of their difference can be described directly in terms of the original estimation error. 
\subsubsection{Mismatched Case} \label{sec: scalar mismatch}
We now turn to the scenario of mismatched estimation, where the underlying
clean signal $X$ is distributed according to $P$, while the decoder believes the
law to be $Q$. \cite{ver_mis10} presents the following relationship between the
relative entropy of the true and mismatched output laws, and the difference
between the mismatched and matched estimation losses:
\bea
D( P \ast \Nscr(0,1/\tsnr) || Q \ast \Nscr(0,1/\tsnr) ) = \frac{1}{2} \int_0^{\tsnr} mse_{P,Q}(\gamma) - mse_{P,P}(\gamma) \,d\gamma, \label{eq: verdu scalar mismatch}
\eea
where $\ast$ denotes the convolution operation, and $mse_{P,Q}(\gamma)$ is defined as 
\bea
\text{mse}_{P,Q}(\gamma) = \E_{P} [(X - \E_{Q}[X | Y_\gamma])^2]. \label{eq: mismatched mse}
\eea     
Towards deriving a pointwise extension of (\ref{eq: verdu scalar mismatch}), we note that it can be recast, assuming again the observation model in (\ref{eq: Brownian motion coupling}),  as the expectation identity
\bea
\E \big[ \log \frac{d P_{Y_\tsnr}}{d Q_{Y_\tsnr}} \big] = \E\bigg[\frac{1}{2}\int_0^\tsnr (X - \E_{Q}[X | Y_\gamma])^2 - (X - \E_{P}[X | Y_\gamma])^2 \,d\gamma \bigg]. \label{eq: verdu mismatch as expectation}
\eea
Let $Z_M$ denote the difference between the random quantities appearing in the above expression, i.e.
\bea 
Z_M = \log \frac{d P_{Y_\tsnr}}{d Q_{Y_\tsnr}} - \frac{1}{2}\int_0^\tsnr (X - \E_{Q}[X | Y_\gamma])^2 - (X - \E_{P}[X | Y_\gamma])^2 \,d\gamma. \label{eq: define Z_M}
\eea
In the following, we provide an explicit characterization of this random variable.
\begin{proposition} \label{lemma: pointwise scalar mismatch}
Assuming $X$ has finite variance under both $P$ and $Q$, $Z_M$ defined in (\ref{eq: define Z_M}), satisfies
\bea
Z_M = \int_0^\tsnr (\E_{P}[X | Y_\gamma] - \E_{Q}[X | Y_\gamma]) \cdot \,dW_\gamma \ \ \ P-a.s.
\eea
\end{proposition}
We observe that the above It\^{o} integral is a martingale and consequently has zero expectation $\E[Z_M]=0$, recovering (\ref{eq: verdu mismatch as expectation}), i.e.Verd\'u's relation from \cite{ver_mis10}. A further implication that can be  read off of Proposition \ref{lemma: pointwise scalar mismatch} rather immediately (as will be explicitly shown in the Section \ref{sec: proofs}) is the following:  
\begin{theorem} \label{thm: variance Z_M scalar mismatched case} 
Assuming $X$ has finite variance under both $P$ and $Q$, $Z_M$ defined in (\ref{eq: define Z_M}), satisfies
\bea 
\text{Var}(Z_M) = \int_0^{\tsnr} mse_{P,Q}(\gamma) - mse_{P,P}(\gamma) \,d\gamma = 2 D( P \ast \Nscr(0,1/\tsnr) || Q \ast \Nscr(0,1/\tsnr) ). \label{eq: variance Z_M scalar mismatched case}
\eea
\end{theorem}
Similarly as in the non-mismatched case, we observe that the variance of the difference between the information and estimation theoretic random variables whose expectations comprise the respective two sides of Verd\'u's mismatch relationship  has a distribution independent characterization in terms of the matched and mismatched estimation errors and consequently, by yet another application of this same relationship of Verd\'u,  in terms of the relative entropy between the output distributions. In the following subsection we extend this line of inquiry and results from the scalar case to that where the channel input is a continuous-time process. 
\subsection{Continuous Time} \label{sec: continuous time}
We now turn to the continuous-time Gaussian channel. Let $X_0^T$ be the underlying noise-free process (with finite power) to be estimated. The continuous time channel is characterized by the following relationship between the input and output processes, 
\begin{equation}
dY_t = X_t \,dt + \,dW_t, \label{eq: channel}
\end{equation}
where $\{W_t\}_{t \geq 0}$ is a standard Brownian motion, independent of $X_0^T$.
\subsubsection{``Pointwise Duncan''} \label{sec: pointwise duncan}
In \cite{duncan}, Duncan proved the equivalence of input-output mutual information to the filtering squared error, of a finite powered continuous time signal $X_t$, corrupted according to (\ref{eq: channel}) to yield the noise corrupted process $Y_t$. The signal is observed for a time duration $[0,T]$. Denoting the time averaged filtering squared error,
\bea
\text{cmmse}(T) &=& \int_0^T E [(X_t - E[X_t | Y^t])^2] dt \label{eq: define cmmse(T)}
\eea
and letting $I(X^T ; Y^T)$ denote the input-output mutual information,  
Duncan's theorem then tells us that, 
\begin{equation}
I(X^T ; Y^T) = \frac{1}{2} \mbox{$\sf{cmmse}$} (T). \label{eq: duncans result}
\end{equation}
In \cite{kadota71}, Kadota et al. extend this result to communication over channels with feedback, and in the recent \cite{WeissmanKimPermuter11} this result is extended to more general scenarios involving the presence of feedback, and it is shown that (\ref{eq: duncans result}) remains true in these more general cases upon replacing the mutual information on the left hand side with directed information. In \cite{zakai05}, several properties of likelihood ratios and their relationships with estimation error are studied. We now proceed to describe a pointwise characterization of Duncan's theorem. 

Considering the random variable $D(T)$ defined as  
\begin{equation}
D(T) = \log \frac{d P_{Y^T|X^T}}{d P_{Y^T}} - \frac{1}{2} \int_0^T (X_t - E[X_t | Y^t])^2 dt ,  \label{eq: duncan D(T)}
\end{equation}
Duncan's theorem is equivalently  expressed as 
\begin{equation}
E [ D(T)] = 0. 
\end{equation}
We now present an explicit formula for $D(T)$ in the following Proposition.
\begin{proposition} \label{lemma: pointwise duncan using girsanov theory}
Let $D(T)$ be as defined in (\ref{eq: duncan D(T)}). Then,
\bea  \label{eq: stochastic integral for Duncan} 
D(T) =  \int_0^T (X_t - E[X_t | Y^t]) dW_t \ \ \ a.s.. 
\eea
\end{proposition}
Note that on the the right side of (\ref{eq: stochastic integral for Duncan}) is a stochastic integral with respect to the Brownian motion $W_{\cdot}$ driving the noise in the channel. 
With this representation, Duncan's theorem follows from the mere fact that this stochastic integral is a martingale  and, in particular, has zero expectation. 

On applying another basic property of the stochastic integral we get the following interesting result for the variance of $D(T)$. 
\begin{theorem} \label{thm: variance D(T)}
For a continuous-time signal $X_0^T$ with finite power, $D(T)$ as defined in (\ref{eq: duncan D(T)}) satisfies 
\begin{equation} \label{eq: variance D(T)}
Var (D(T)) =  \mbox{$\sf{cmmse}$} (T).
\end{equation}
\end{theorem}
In conjunction with Duncan's theorem (\ref{eq: duncans result}), we get the
following relationship,
\begin{equation}
Var (D(T))  = \mbox{$\sf{cmmse}$} (T) = 2 I(X^T; Y^T) \label{eq: variance-cmmse-mutual information equality}
\end{equation}
which parallels our discovery for scalar random variables, in (\ref{eq: variance
of Z scalar for Brownian motion coupling}). Thus, we find that the pointwise
tracking error satisfies this intriguing distribution independent property, for
the full generality of continuous time inputs for the Gaussian Channel. That the
estimation error and mutual information emerge from this analysis in such a
crisp manner is quite satisfying.  
\begin{remark}
One can note that for $X_t \equiv X$ in the
interval $[0,T]$, we can use the results in (\ref{eq: variance D(T)}) and (\ref{eq: stochastic integral for Duncan}), to recover Theorem \ref{thm: variance of Z scalar for Brownian motion coupling} and its
its pointwise characterization in Proposition \ref{lemma: pointwise scalar Z for
Brownian motion coupling} respectively.
\end{remark}
Among the additional immediate benefits the characterization in (\ref{eq: stochastic integral for Duncan}), is that it allows us to infer facts about the limiting behavior of the random variables involved, such as in the following theorem.
\begin{theorem} \label{thm: limit theorem}
Suppose that the process $\{X_t\}_{t \geq 0}$ satisfies
\begin{equation} \label{eq: condition on growth rate}
\lim_{T \rightarrow \infty} \frac{1}{T^2} \mbox{$\sf{cmmse}$} (T) = 0
\end{equation}
(or, equivalently, by Duncan's theorem, $\lim_{T \rightarrow \infty} I(X^T; Y^T)/T^2 = 0$	). Then, 
\bea
\text{l.i.m.}_{T \rightarrow \infty} \frac{1}{T} \left[  \log \frac{dP_{Y_0^T|X_0^T}}{dP_{Y_0^T}} - \frac{1}{2} \int_0^T (X_t - \E[X_t|Y_0^t])^2 \,dt \right]  = 0,  \label{eq: limit theorem}
\eea
where $\text{\text{l.i.m.}}$ denotes Limit in the Mean. 
\end{theorem} 
We already know from Duncan's theorem that the two quantities that make up
$D(T)$, namely the information density and the causal estimation error,  are
equal in expectation for every $T>0$, but our formulation reveals much more
about the pointwise behavior of these random quantities in themselves. In
particular, it is interesting to note that so little is needed to guarantee the
convergence in (\ref{eq: limit theorem}): not even wide sense stationarity of
the marginal of the underlying process is required. Indeed, any process with
$\tvar(X_t)$ growing sublinearly with $t$ is easily seen to satisfy (\ref{eq:
condition on growth rate}).
  
\subsubsection{Pointwise Mismatch} \label{sec: pointwise mismatch}
We now consider the setting in \cite{weissman10}, where a continuous time signal $X_t$, distributed according to a law $P$ is observed through additive Gaussian noise, and is estimated by an estimator that would have been optimal if the signal had followed the law $Q$. In this general setting , the main result in \cite{weissman10} shows that the relative entropy between the laws of the output for the two different underlying distributions (P and Q), is exactly half the difference between the mismatched and matched filtering errors.  
Let $Y_t$ be the continuous time AWGN corrupted version of $X_t$ as given by (\ref{eq: channel}).  Let $P_{Y_0^T}$ and $Q_{Y_0^T}$ be the output distributions when the underlying signal $X_0^T$ has law $P$ and $Q$, respectively. As before, T denotes the time duration for which the process is observed. We denote the mismatched causal mean squared error, 
\bea
cmse_{P,Q}(T) = \int_0^TE_P[(X_t - E_Q[X_t | Y^t])^2] \,dt.
\eea
In this setting, \cite{weissman10} tells us that the relative entropy between the output distributions is half the difference between the mismatched and matched filtering errors, i.e. 
\begin{equation}
D(P_{Y_0^T}||Q_{Y_0^T}) = \frac{1}{2} [cmse_{P,Q}(T) - cmse_{P,P}(T)]. \label{eq: mismatched result}
\end{equation}
Define the pointwise difference between the log Radon-Nikodym derivative and half the mismatched causal squared error difference,
\begin{equation}
M(T) = \log \frac{d P_{Y_0^T}}{d Q_{Y_0^T}} - \frac{1}{2}\int_0^T\bigg[(\E_Q[X_t | Y^t] - X_t)^2 - (\E_P[X_t | Y^t] - X_t)^2\bigg]\,dt. \label{eq: define M(T)}
\end{equation}
Note that according to the above definition, (\ref{eq: mismatched result}) can be equivalently stated as 
\bea
\E[M(T)] = 0. 
\eea
But in fact much more can be said about $M(T)$: 
\begin{proposition} \label{lemma: mismatch M(T)}
\begin{equation}
M(T) = \int_0^T(\E_P[X_t | Y^t] - \E_Q[X_t | Y^t]) \,dW_t  \ \ \ P-a.s.,
\end{equation} 
where $M(T)$ is as defined in (\ref{eq: define M(T)}), and $X_t$ is assumed to have finite power under the laws $P$ as well as $Q$.
\end{proposition}
We note that relation (\ref{eq: mismatched result}) is implied immediately by Proposition \ref{lemma: mismatch M(T)} due to the `zero mean' property of the martingale $M(T)$. But more can be read off of this result. For example, the following characterization of the variance of $M(T)$ will be shown in Section \ref{sec: proofs} to follow quite directly from Proposition \ref{lemma: mismatch M(T)}.  
\begin{theorem} \label{thm: variance M(T)}
$M(T)$ as defined in (\ref{eq: mismatched result}), satisfies 
\begin{equation}
Var(M(T)) = cmse_{P,Q}(T) - cmse_{P,P}(T) = 2 D(P_{Y_0^T}||Q_{Y_0^T}).
\end{equation}
\end{theorem}
Thus, the variance of $M(T)$ is exactly the difference between the causal mismatched and matched squared errors. And further, from \cite{weissman10} we know that it is equal to twice the relative entropy between the output distributions according to laws $P$ and $Q$. 

\subsubsection{Presence of Feedback} \label{sec: presence of feedback}
In the previous subsections, we explicitly characterized a pointwise
relationship between the log Radon-Nikodym derivates associated with the
informational quantities and the squared filtering error. These
characterizations give us a crisp understanding of well known
information-estimation results such as Duncan's theorem \cite{duncan}, and the
equivalence between mismatched estimation and relative entropy
\cite{weissman10} for the continuous time setting. These results emerge as
direct corollaries of our characterization of the tracking errors as stochastic
integrals. Further, we establish a new equivalence between estimation error and
variance of the tracking error.

In this section we revisit \cite{kadota71}, where Kadota et al. present a generalization of Duncan's theorem for the additive Gaussian channel with feedback.  The channel input $\phi_t$ is a function of the underlying process $X_t$ as well as the past outputs of the channel $Y_0^t$, in an additive Gaussian noise setting. The observation window is $t \in [0,T]$. The channel can be represented as,
\begin{equation}
Y_t = \int_0^t \phi_s(Y_0 ^{s} ,X_s) ds + W_t, \label{general channel} 
\end{equation}
where, as usual, the standard Brownian motion $W_{\cdot}$ is independent of the underlying process $X_{(\cdot)}$. In differential form, (and using shorthand to represent $\phi_t(Y_0 ^{t} ,X_t)$) we can rewrite (\ref{general channel}) as
\begin{equation}
dY_t = \phi_t \,dt + \,dW_t  
\end{equation}
We denote the causal estimate of $\phi_t$ based on observations up until $t$ by
\begin{equation}
\hat{\phi}_t = E [ \phi_t | Y_0^t]. \label{estimate of phi} 
\end{equation}
Under mild regularity conditions on $\phi_t$, the mutual information between the
input and output is equal to half the causal mean squared error.  With our
notation, the main result of \cite{kadota71} is expressed as
\begin{equation}
I(X_0^T;Y_0^T) = \frac{1}{2} \int_0^T E[ (\phi_t - \hat{\phi}_t)^2] \,dt. \label{main result kadota}
\end{equation}
We use Girsanov theory to develop a pointwise relationship in this setting, akin to our treatment of Duncan's theorem.
Defining
\bea
D_{\phi}(T) &\stackrel{\Delta}{=}& \log \frac{d P_{Y^T|X^T}}{d P_{Y^T}} - \frac{1}{2} \int_0^T (\phi_t - \hat{\phi}_t)^2 \,dt, \label{eq: define D phi}
\eea
We have the following:
\begin{theorem} \label{thm: pointwise result for phi channel}
For $\phi_t$ which satisfy a finite power criterion, we have
\begin{equation}
D_{\phi}(T) = \int_0^T(\phi_t - \hat{\phi}_t) \,dW_t \ \ \ a.s., \label{eq: pointwise result for phi channel}
\end{equation}
where $D_{\phi}(T)$ is as defined in (\ref{eq: define D phi}).
\end{theorem}

Parallel to our discovery in the pointwise treatment of Duncan's theorem, we can use Theorem \ref{thm: pointwise result for phi channel} to deduce various results. Note from (\ref{eq: pointwise result for phi channel}), that $D_{\phi}(T)$ is a martingale. Therefore,
\bea
\E[D_{\phi}(T)] = 0,
\eea
recovering the main result of \cite{kadota71}, namely (\ref{main result kadota}). Using It\^{o}'s Isometry we also immediately obtain:
\begin{corollary} \label{corollary: second moment phi channel}
\begin{eqnarray}
Var(D_{\phi}(T)) &=& \int_0^T E[ (\phi_t - \hat{\phi}_t)^2 ] \,dt \\
 &=& \mbox{$\sf{cmmse}$}_{\phi}(T)
\end{eqnarray}
\end{corollary}
The above follows directly from the application of It\^{o}'s Isometry property to the stochastic integral in (\ref{eq: pointwise result for phi channel}) and noting that $\E[D_{\phi}(T)] = 0$.

Thus, even for the generalized setting of communication over channels with feedback, we can characterize how closely the information density and squared filtering error track each other. We note that the second moment of the tracking error is equal to the filtering error for all finite powered distributions of the underlying signal. In particular, these results may have applications in approximating the mutual information via estimation theoretic quantities, for channels with feedback. In the special case when $\phi_t = X_t$, we recover the results obtained in the pointwise treatment of Duncan's theorem in Section \ref{sec: pointwise duncan}. Here, we would like to note that Theorem \ref{thm: pointwise result for phi channel} can further be extended to accommodate mismatch. 

Let us denote separately, the Causal Estimates of $\phi_t$ under the two laws $P$ and $Q$ that govern the underlying process $X_t$: 
\bea
\hat{\phi}^P_t = \E_P[\phi_t | Y_0^t] \\
\hat{\phi}^Q_t = \E_Q[\phi_t | Y_0^t]
\eea 
Define,
\bea
M_{\phi}(T) \stackrel{\Delta}{=} \log \frac{d P_{Y_0^T}}{d Q_{Y_0^T}} - \frac{1}{2}\int_0^T\bigg[(\hat{\phi}^Q_t - \phi_t)^2 - (\hat{\phi}^P_t - \phi_t)^2\bigg]\,dt. \label{eq: define M phi}
\eea
As will be shown in the next section, Girsanov theory allows us to establish, for this setting
\bea
D(P_{Y_0^T}||Q_{Y_0^T}) = \frac{1}{2} [cmse^{\phi}_{P,Q}(T) - cmse^{\phi}_{P,P}(T)], \label{eq: D-mse for phi channel}
\eea
where
\bea
cmse^{\phi}_{P,Q}(T) = \int_0^T \E_P[(\phi_t - \hat{\phi}^Q_t)^2] \,dt.
\eea
Define,
\bea
M_{\phi}(T) = \log \frac{d P_{Y_0^T}}{d Q_{Y_0^T}} - \frac{1}{2}\int_0^T\big[(\hat{\phi}^Q_t - \phi_t)^2 - (\hat{\phi}^P_t - \phi_t)^2\big]\,dt. \label{eq: define M_{phi}(T)}
\eea
The arguments given in the proof of Proposition \ref{lemma: mismatch M(T)}, and the treatment of the non-mismatched case above, can be carried over to show the following:
\begin{theorem} \label{thm: mismatch general channel}
For a finite power constraint on $\phi_t$ under laws $P$ and $Q$, $M_{\phi}(T)$ as defined in (\ref{eq: define M_{phi}(T)}) satisfies
\bea
M_{\phi}(T) = \int_0^T(\hat{\phi}^P_t - \hat{\phi}^Q_t) \,dW_t \ \ \ P-a.s.
\eea
\end{theorem}
Note that (\ref{eq: D-mse for phi channel}) follows directly from Theorem \ref{thm: mismatch general channel}, by noting that $M_{\phi}(T)$ is a martingale and consequently has zero mean. Thus, the generalized D-MSE relationship for channels with feedback is an expectation identity that arises from the pointwise treatment of the tracking error in (\ref{eq: define M_{phi}(T)}).
As a corollary, we also obtain the generalized result for the second moment of the tracking error, which acts as a bridge between the relative entropy and the difference of the mismatched and matched filtering errors,
\begin{corollary} \label{corollary: second moment phi channel with mismatch}
\bea
\text{Var}(M_{\phi}(T)) = cmse^{\phi}_{P,Q}(T) - cmse^{\phi}_{P,P}(T) = 2 D(P_{Y_0^T}||Q_{Y_0^T}) \label{eq: mismatch with phi}
\eea
\end{corollary}
The result in (\ref{eq: mismatch with phi}) can be specialized to obtain results similar to those obtained previously for ``Pointwise Duncan'', and pointwise mismatch in the absence of feedback.  
\subsubsection{Pointwise I-MMSE for processes} \label{sec: pointwise i-mmse for processes}
In \cite{gsv2005}, Guo et al. present what is known as the I-MMSE relationship
for the Gaussian channel. This result states that the mutual information (at
signal-to-noise ratio level `snr') is the integral over SNR (from level 0 to
`snr') of half the non-causal squared error.
 
In this subsection, we present a characterization of the pointwise nature of the I-MMSE relationship for processes in the continuous time Gaussian channel. We first explain the channel model. Since we are now concerned with two continuously varying parameters, namely time and `snr', the Gaussian noise corrupting the signal is a standard Brownian sheet $W_{t,\gamma}$. For a fixed $\gamma$, we let $W_{\cdot}^{(\gamma)}$ denote the Brownian motion defined by $W^{(\gamma)}_t = W_{t, \gamma}$. The channel then, at SNR $\gamma$,  is given by
\bea
dY^{(\gamma)}_t = \gamma X_t\,dt + dW^{(\gamma)}_t, \label{eq: time-snr continuous channel}
\eea
where $\{X_t\}_{0 \leq t \leq T}$ is the underlying noise free process, which is
independent of the Brownian sheet. The output of the channel at SNR $\gamma$, is
denoted by $Y_0^{T,(\gamma)} = \{ Y_t^{(\gamma)}\}_{0 \leq t \leq T}$.
In this framework, the I-MMSE relationship from \cite{gsv2005} tells us that,
\bea 
I(X_0^T;Y_0^{T,(\tsnr)}) = \frac{1}{2} \int_0^\tsnr \text{mmse}(\gamma) \,d\gamma, \label{eq: I-MMSE for processes 1}
\eea
where
\bea
\text{mmse}(\gamma) = \int_0^T \E [ ({X}_t - \E[{X}_t | Y_0^{T,(\gamma)}])^2 ] \,dt. \label{eq: define mmse for processes}
\eea
Note that (\ref{eq: I-MMSE for processes 1}) can now equivalently be stated as ,
\bea 
\E\big[\log \frac{dP_{{Y}_0^{T,(\tsnr)} |{X}_0^T}}{dP_{{Y}_0^{T,(\tsnr)}}}\big] =  \E \big[\frac{1}{2} \int_0^\tsnr \int_0^T ({X}_t - \E[{X}_t | Y_0^{T,(\gamma)}])^2 \,dt \,d\gamma\big]. \label{eq: I-MMSE relationship for processes}
\eea
Defining the pointwise difference between the input-output information density and half the non-causal error integrated over time and SNR
 \bea
 N(T) \stackrel{\Delta}{=} \log \frac{dP_{{Y}_0^{T,(\gamma)} |{X}_0^T}}{dP_{{Y}_0^{T,(\gamma)}}} -  \frac{1}{2} \int_0^\tsnr \int_0^T ({X}_t - \E[{X}_t | Y_0^{T,(\gamma)}])^2 \,dt \,d\gamma, \label{eq: define N(T)}
 \eea
we first note that the I-MMSE relationship  in (\ref{eq: I-MMSE relationship for processes}) can equivalently be stated as 
 \bea
 \E[N(T)] = 0.
 \eea
In doing so, we formulate the I-MMSE relationship as an expectation identity. In the previous discussions, we observed an estimation theoretic flavor to the second moment of the tracking error in characterizations of both Duncan's result and that of mismatched estimation. This kind of a relationship turns out to hold also in our present context: 
 \begin{theorem} \label{thm: pointwise I-MMSE processes}
For a finite power continuous-time process $X^T$, $N(T)$ as defined in (\ref{eq: define N(T)}) satisfies
 \bea
 Var(N(T)) = \int_0^{\tsnr} \text{mmse}(\gamma) \,d \gamma = 2 I(X_0^T;Y_0^{T,(\tsnr)}).
 \eea
 \end{theorem}
In establishing the above result in the Section \ref{sec: proofs}, we use a
 multidimensional version of Girsanov's theorem to characterize the input-output
information density of a piecewise constant process observed through AWGN. We
then use approximation arguments to extend the result to the class of finite
power continuous time processes.
\subsubsection{Pointwise causal vs. non-causal error} \label{sec: pointwise causal vs. non-causal error}
By a combination of Duncan's theorem and the I-MMSE result, the authors of \cite{gsv2005} establish the equivalence of the causal error at SNR level `snr', and the non-causal error averaged over SNR uniformly distributed between 0 and `snr'. The input-output mutual information acts as a bridge between the quantities. Let `$\text{cmmse}(\tsnr)$' denote the integral of the filtering error for the channel described in (\ref{eq: time-snr continuous channel}),
\bea 
\text{cmmse}(\tsnr) = \int_0^T \E [ ({X}_t - \E[{X}_t | \{Y^{(\tsnr)}_s\}_{0 \leq s \leq t}])^2 ] \,dt. \label{eq: define cmmse for processes}
\eea
Recalling the definition of the non-causal error `$\text{mmse}(\gamma)$' in (\ref{eq: define mmse for processes}), the causal vs. non-causal error relationship is
\bea 
\text{cmmse}(\tsnr) = \frac{1}{\tsnr} \int_0^\tsnr \text{mmse}(\gamma) \,d\gamma. \label{eq: causal vs. non-causal relationship}
\eea
So far, we have presented pointwise characterizations of Duncan's result in Section \ref{sec: pointwise duncan} as well as the I-MMSE relationship  for continuous time processes in Section \ref{sec: pointwise i-mmse for processes}. Using these two characterizations, in Section \ref{sec: proofs} we develop a pointwise version of the celebrated estimation-theoretic result (\ref{eq: causal vs. non-causal relationship}). Specifically, we show that under the Brownian sheet induced channel described in (\ref{eq: time-snr continuous channel}), and under some regularity assumed on the process $X^T$, the difference 
\begin{equation} \label{eq: pointwise difference between causal and non-causal}
\int_0^T (X_t - E[X_t | Y_0^{t, (snr)}])^2 dt - \frac{1}{snr} \int_0^{snr}  \left[  \int_0^T (X_t - E[X_t | Y_0^{T, (\gamma)}])^2 dt  \right] d \gamma
\end{equation}
can be characterized as a difference between stochastic integrals. In particular, such a characterization immediately implies (\ref{eq: causal vs. non-causal relationship}). 

\subsubsection{Pointwise causal vs. anticausal error}
Now, we present another interesting application of Proposition \ref{lemma: pointwise duncan using girsanov theory}, a pointwise treatment of the causal vs. anti-causal estimation error relationship. Duncan's theorem gives us the remarkable equality between the causal squared error and input-output mutual information for the continuous-time Gaussian channel. Invoking the invariance of mutual information to the direction of time, it can be observed (as is noted in \cite{gsv2005}) that the causal squared error is equal to the anticausal squared error (for a given `snr'), regardless of the input distribution of the underlying process.  
Let $X_t$ be the noise free stochastic process that is distributed according to law P. Let $Y_t$ be the continuous time AWGN corrupted version of $X_t$ at $snr=1$, according to (\ref{eq: channel}). Let the observation window be $t \in [0,T]$.

Let us now denote 
\bea \label{eq: tilded process for causal anticausal error beg}
\tilde{X}_t &=& X_{T-t}, \\
\tilde{Y}_t &=& Y_T - Y_{T-t}, \\
B_t &=& W_T - W_{T-t}. \label{eq: tilded process for causal anticausal end}
\eea   
Note that the anti-causal estimation error for the original processes $(X^T, Y^T)$ is given by the the causal estimation error associated with these ``tilded'' processes $(\tilde{X}^T, \tilde{Y}^T)$, i.e. : 
\bea \label{eq: anti-causal error}
\int_0^T E \big[ ( \tilde{X}_t - \E[ \tilde{X}_t| \tilde{Y}_0^t])^2 \big] \,dt &=&  \int_0^T E \big[ ( {X}_t - \E[ {X}_t | {Y}_t^T])^2 \big] \,dt.
\eea

Define the difference of the causal and anti-causal estimation errors,
\bea
J(T) \stackrel{\Delta}{=} \int_0^T (X_t - \E[X_t|Y_0^t])^2 \,dt -  \int_0^T ( \tilde{X}_t - \E[ \tilde{X}_t| \tilde{Y}_0^t])^2 \,dt \label{eq: define J(T)}
\eea
Then, we have the following pointwise result relating the difference of the causal and anti-causal errors
\begin{proposition} \label{lemma: difference of causal anti-causal errors}
For a process $X^T$ with finite power, $J(T)$ as defined in (\ref{eq: define J(T)}) satisfies
\bea \label{eq: pointwise difference of errors} 
\frac{1}{2} J(T)   &=&  \int_0^T  E[X_t|Y_0^t] \cdot dW_t -  \int_0^T  E[ \tilde{X}_t| \tilde{Y}_0^t] \cdot dB_t  \ \ \ a.s.
\eea
\end{proposition}
We note that the right hand side of equation (\ref{eq: pointwise difference of errors}) is the difference of two martingales. Taking expectation on both sides, we recover the equality of causal and anti-causal squared error,
\bea
\int_0^T \E \big[ (X_t - \E[X_t|Y_0^t])^2 \,dt \big] &=& \int_0^T \E \big[ ( \tilde{X}_t - \E[ \tilde{X}_t| \tilde{Y}_0^t])^2\big]  \,dt, \label{eq: causal anti-causal}
\eea
using merely the fact that the It\^{o} integrals on the right hand side of (\ref{eq: pointwise difference of errors}) have zero mean.
Thus, we provide a pointwise characterization of the causal vs. anti-causal errors. Duncan's theorem implies, what is otherwise a surprising result, that the causal and anti-causal squared errors are equal, as rederived in (\ref{eq: causal anti-causal}). Through (\ref{eq: pointwise difference of errors}), we uncover the structure of and dependence between the random quantities involved and characterize their difference as a difference of two zero mean stochastic integrals.
\section{Scalar Setting: Examples, alternative Couplings, further Observations, and Identities } \label{sec: revisiting the scalar setting}
 Returning to the scalar channel setting of Subsection \ref{sec: scalar estimation}, we introduce notation as follows: 
\begin{define} \label{def: scalar setting definitions}
\bea
I_1 &=& \log\frac{dP_{Y_{\tsnr}|X}}{dP_{Y_{\tsnr}}} \label{I_1}\\
I_2 &=& \frac{1}{2}\int_0^{\tsnr} (X - \E[X|Y_{\gamma}])^{2} \,d\gamma \label{I_2}\\
Z &=& I_1 - I_2, \label{Z}
\eea
where $Z$, as in the previous section, is informally referred to as the ``tracking error'' 
\end{define}
\subsection{The Original Coupling} \label{sec: scalar example 1}
We studied the example of the scalar Gaussian channel corrupted by additive Gaussian noise where the additive noise components for the different SNR levels were coupled via a standard Brownian motion, as in (\ref{eq: Brownian motion coupling}). We characterized explicitly in Proposition \ref{lemma: pointwise scalar Z for Brownian motion coupling}, the tracking error $Z$ between the information density and half the estimation error. 

To illustrate how explicit this characterization allows us to be, let us consider the case of  $X \sim \Nscr(0,1)$, and use Proposition \ref{lemma: pointwise scalar Z for Brownian motion coupling}  to characterize the distribution of the random variable $Z$. Note that in this case,
\bea
\E[X|Y_\gamma] = \frac{\gamma X + W_\gamma}{\gamma + 1}. \label{eq: Z scalar for Brownian motion coupling with Gaussian input}
\eea 
Combining with (\ref{eq: pointwise scalar Z for Brownian motion coupling}), we have
\bea
Z = \int_0^\tsnr \frac{X + W_\gamma}{\gamma + 1} \,dW_\gamma \ \ \ a.s.
\eea 
Also, invoking the result in Theorem \ref{thm: variance of Z scalar for Brownian motion coupling}, we obtain the variance of $Z$ to be
\bea 
\text{Var}(Z) &=& 2 I(X;Y_\tsnr) \\
&=& \log (1 + \tsnr). \label{eq: Z scalar variance for X Gaussian in Brownian motion coupling}
\eea 

We shall now look at the pointwise scalar estimation problem in a new light. Recall that in moving from (\ref{eq: i-mmse form 1}) to (\ref{eq: exchange expectation and integral in I-MMSE scalar}), we place all the random variables $(X,Y_0^\tsnr)$ on the same probability space, via a standard Brownian motion, as in (\ref{eq: Brownian motion coupling}). Note, however, that the only assumption for the original results that hold in expectation is that, for each $\gamma > 0$, the channel satisfies (\ref{eq: gaussian channel condition}), i.e. 
\bea
Y_{\gamma}|X \sim \Nscr(\sqrt{\gamma} X, 1), \label{eq: channel_cond}
\eea
where $\Nscr(\mu,\sigma^2)$ denotes the Gaussian distribution with mean $\mu$ and variance $\sigma^2$. Taking the channel noise variables for the various SNR levels to be the components of a Brownian motion, as in (\ref{eq: Brownian motion coupling}), is but one possibility for a coupling that respects (\ref{eq: channel_cond}). 

For $Z$ as defined in (\ref{Z}), the I-MMSE relationship tells us that $\E[Z]=0$, for all such $(X,Y_0^\tsnr)$ that are consistent with (\ref{eq: channel_cond}). It is instructive to note that (\ref{eq: channel_cond}) is a requirement on the channel for the individual SNR levels. As mentioned, however, there are several ways in which we can couple the input $X$ and outputs $\{Y_0^\tsnr \}$ together so that they satisfy (\ref{eq: channel_cond}). The I-MMSE relationship implies that for all such couplings we have $\E[Z] = 0$. Before exploring some other examples of such `couplings' and their properties, let us note a refinement of this zero-mean property pertaining to the random variable $Z$, which holds regardless of the coupling.
\begin{proposition} \label{lemma: conditional expectation of Z scalar}
Suppose $X$ has finite variance and that $Z$ is defined as in Definition \ref{def: scalar setting definitions}, under a  joint distribution on $(X, Y_0^{snr})$  satisfying (\ref{eq: channel_cond}). Then,
\bea
\E[Z | X] = 0, \ \ \ a.s. \label{eq: conditional expectation of scalar Z}
\eea 
\end{proposition}
Thus, not only is the tracking error a zero-mean random variable, but even its
conditional expectation $\E[Z|X]$ is zero. We use the setting and results in
\cite{ver_mis10} to establish this result. The I-MMSE relationship which states
that $\E[Z]$ = 0, is then immediately implied by Proposition \ref{lemma:
conditional expectation of Z scalar}. 

Having briefly touched upon the idea of ways other than the channel in (\ref{eq: Brownian motion coupling}), in which we can comply with the marginal channel requirements in (\ref{eq: channel_cond}), let us look at some concrete examples and draw a comparison between them.

\subsection{Additive Standard Gaussian} \label{sec: scalar example 2}
\begin{figure}
\begin{center}
\includegraphics[height=4in,width=6in]{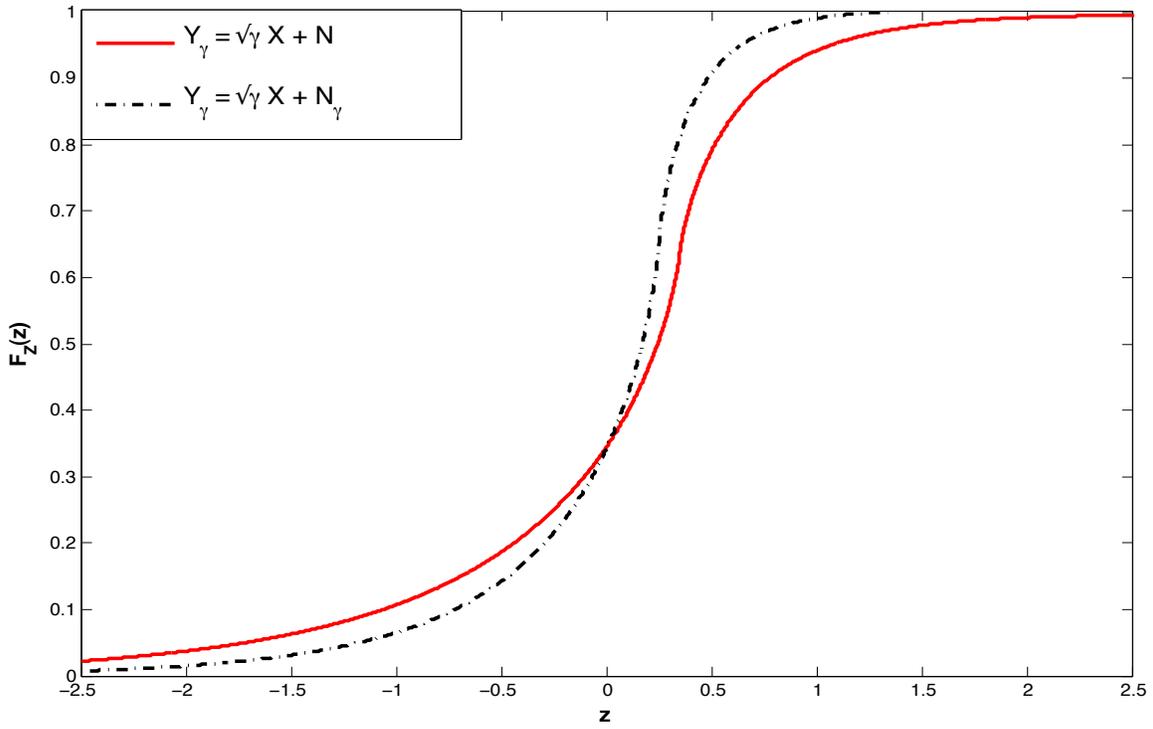}
\caption{C.D.F. of $Z$ under the couplings discussed in Subsections \ref{sec: scalar example 2} and \ref{sec: scalar example 3} for a Gaussian input}
\label{cdf}
\end{center}
\end{figure}

An alternative coupling between $X$ and $Y_\gamma$ that respects (\ref{eq: channel_cond}), is achieved by using a scaled standard Gaussian random variable as additive noise, instead of the Brownian motion considered in the previous setting. The channel is described by letting, for $\gamma \in [0,\text{\tsnr}]$,
\bea
Y_\gamma = \sqrt{\gamma}X + N, \label{eq: channel in example 1}
\eea 
where $N \sim \Nscr(0,1)$ is independent of $X$. Note that in this coupling, the channel has the same noise component for all values of $\gamma$. We now present the pointwise characterization of the `tracking error' for this setting in the following Lemma.
\begin{lemma} \label{lemma: Z scalar additive standard Gaussian}
Let $X \sim P_X$, have a finite second moment. For the channel in (\ref{eq: channel in example 1}), the pointwise tracking error $Z$ defined in Definition \ref{def: scalar setting definitions} can be expressed as,
\bea
Z = \int_0^{\tsnr} \tilde{Z}_\gamma \,d\gamma, \label{eq: Z scalar for additive standard Gaussian coupling}
\eea
where $\tilde{Z}_\gamma$ is given by 
\bea
\tilde{Z}_\gamma &=& \frac{1}{2}\big{\{}\E[X^2|Y_\gamma] - X\E[X|Y_\gamma] - (X
- \E[X|Y_\gamma])^2 + \frac{Y_\gamma}{\sqrt{\gamma}}(X -
\E[X|Y_\gamma])\big{\}}. \label{eq: tracking error for additive standard
gaussian channel in differential version of i-mmse}
\eea
\end{lemma}
As a sanity check, one can observe that $\E[\tilde{Z}_\gamma | Y_\gamma] = 0$
and thus $\E[\tilde{Z}_\gamma] = 0$. Consequently, $\E[Z] =
\int_0^{\tsnr}\E[\tilde{Z}_\gamma]\,d\gamma = 0$. Lemma \ref{lemma: Z scalar
additive standard Gaussian} is closely related to the pointwise identity in
\cite[Theorem 2.3]{Guo_Thesis}. However, for completeness we present a
stand-alone proof in Section \ref{sec: proofs}.
\begin{example}
Applying Lemma \ref{lemma: Z scalar additive standard Gaussian} in the case
where  the channel input $X$ is standard normal, i.e., $X \sim \Nscr(0,1)$, the
tracking error is given by
\bea
Z &=& \frac{1}{2}\bigg[\log(1+\tsnr) - N^2\log(1+\tsnr) + 2XN \tan^{-1}(\sqrt{\tsnr}) \bigg] \ \ \ a.s. \label{eq: Z scalar for additive standard Gaussian coupling for X normal}
\eea
and thus, in particular, the variance of the tracking error is 
\bea
\text{Var}(Z) &=& \frac{1}{2}\big(\log(1+ \tsnr)\big)^2 + \big(\tan^{-1}(\sqrt{\tsnr})\big)^2. \label{eq: Z scalar variance for X Gaussian for Additive Standard Gaussian coupling}
\eea
\end{example}
For snr=1, we present a plot of the Cumulative Distribution Function of the random variable $Z$ in (\ref{eq: Z scalar for additive standard Gaussian coupling for X normal}) in Figure \ref{cdf}.
\subsection{Independent Standard Gaussian's} \label{sec: scalar example 3}
We present yet another illustration of a different coupling which places the input and outputs of the Gaussian channel (\ref{eq: channel_cond}) on the same probability space. Unlike the previous two examples in Sections \ref{sec: scalar example 1} and \ref{sec: scalar example 2} respectively, we here look at the limiting behavior of a family of couplings achieved by the construction described below.

Let $\Delta > 0$. Define $\Delta = \frac{\tsnr}{M}$ for $M$ natural. Let
$\Iscr_i \equiv [ (i-1)\Delta, i\Delta )$ for $i \in \{1,2,3 \dots M\}$. Let
$N_i$ be independent standard Gaussian random variables $\sim \Nscr(0,1)$. Now
we define the following process
\bea
Y_\gamma = \sqrt{\gamma}X + N_\gamma,
\eea
where $N_\gamma = N_i$ for $\gamma \in \Iscr_i$. Note that this is a coupling of
the channel noise components at different SNR's that adheres to (\ref{eq:
channel_cond}).

We now evaluate $I_2$ as defined in (\ref{I_2}) for this process
\bea
I_2 &=& \frac{1}{2}\int_0^{\tsnr} (X - \E[X|Y_{\gamma}])^{2}\,d\gamma \\
&=& \sum_{i=1}^{M} \int_{(i-1)\Delta}^{i\Delta} (X - \E[X|Y_{\gamma}])^{2}\,d\gamma \\
\eea
We are interested in the limiting process when $\Delta$ is small. We consider the Riemann sum approximation of the above integral,
\bea
S_M &=& \sum_{i=1}^{M} (X - \E[X|Y_{i\Delta}])^{2} \Delta\\
&=& \frac{\tsnr}{M} \sum_{i=1}^{M} (X - \E[X|Y_{i\Delta}])^{2}
\eea
We now look at the conditional variance of $I_2$ given $X$
\bea
\text{Var}[I_2 | X] = \frac{\tsnr^2}{M^2} \sum_{i=1}^{M} \text{Var}[(X -
\E[X|Y_{i\Delta}])^{2}|X],
\eea
where $\text{Var}[\cdot | X]$ denotes the conditional variance averaged
over the distribution of $X$. Thus, under mild regularity conditions on the
underlying distribution of $X$ (for instance, $\E[X^4] < \infty$ suffices)
we can see that the summand in the r.h.s. is bounded above by a constant
independent of $M$. Therefore the sum involves adding over $M$ terms that
are $O(1)$. In other words
\bea
\text{Var}[I_2 | X] = \frac{O(M)}{M^2} \hspace{1em} \text{in probability},
\eea
and thus
\bea
\text{l.i.m.}_{M \rightarrow \infty} I_2 &=& \E[I_2 | X]. \label{eq: limit M
tending to infinity}
\eea
Therefore, $Z = I_1 - I_2$, the tracking error is given (in the l.i.m. sense as $M \rightarrow \infty$) by
\bea
Z &=& I_1 - I_2 \\
&=& I_1 - \E[I_2 | X] \\
&=& I_1 - \E[I_1 | X] \label{eq: Z scalar for independent standard Gaussians coupling}
\eea
where the last equality follows from equation (\ref{eq: conditional expectation
of scalar Z}). We now consider the case when $X \sim \Nscr(0,1)$ which satisfies
(\ref{eq: limit M tending to infinity}), and thus we can apply (\ref{eq: Z
scalar for independent standard Gaussians coupling}) to explicitly calculate
$Z$.

Note that $I_1(X,Y_\tsnr)$, as defined in Definition \ref{def: scalar setting definitions}, depends only on the joint distribution of $X$ and $Y_\tsnr$, and is therefore the same for all the channel couplings that are consistent with (\ref{eq: channel_cond}), for a fixed input distribution on $X$. Thus, for $X \sim \Nscr (0,1)$, $I_1$ can be computed explicitly using Definition \ref{def: scalar setting definitions} to yield,
\bea 
I_1 &=& \frac{1}{2}\log(1+\tsnr) + \frac{1}{2}\bigg(\frac{Y^2}{1 + \tsnr} - (Y - \sqrt{\tsnr}X)^2 \bigg) \label{eq: I1 in scalar example 3}
\eea
and
\bea
\E[I_1|X] = \frac{1}{2}\log(1+\tsnr) + \frac{1}{2}\bigg(\frac{\tsnr}{1+\tsnr}(X^2 - 1) \bigg). \label{eq: E[I1] in scalar example 3}
\eea 
Let $N = Y_{\tsnr} - \sqrt{\tsnr}X$. Using (\ref{eq: I1 in scalar example 3}) and (\ref{eq: E[I1] in scalar example 3}), we simplify (\ref{eq: Z scalar for independent standard Gaussians coupling}) to get the following closed form expression for $Z$:
\bea
Z &=& \frac{1}{2(1 + \tsnr)}\bigg(-N^2\tsnr + 2XN\sqrt{\tsnr} + \tsnr\bigg).
\eea 
Further, the variance is given by,
\bea
\text{Var}(Z) = \tsnr\frac{1+2\tsnr}{2(1+\tsnr)^2} \label{eq: Z scalar variance for independent standard Gaussians coupling}
\eea 
For snr=1, we present a plot of the Cumulative Distribution Function of the random variable $Z$ in (\ref{eq: Z scalar for independent standard Gaussians coupling}) in Figure \ref{cdf}.
\subsection{Comparison of Variances}
\begin{figure}
\begin{center}
\includegraphics[height=4in,width=6in]{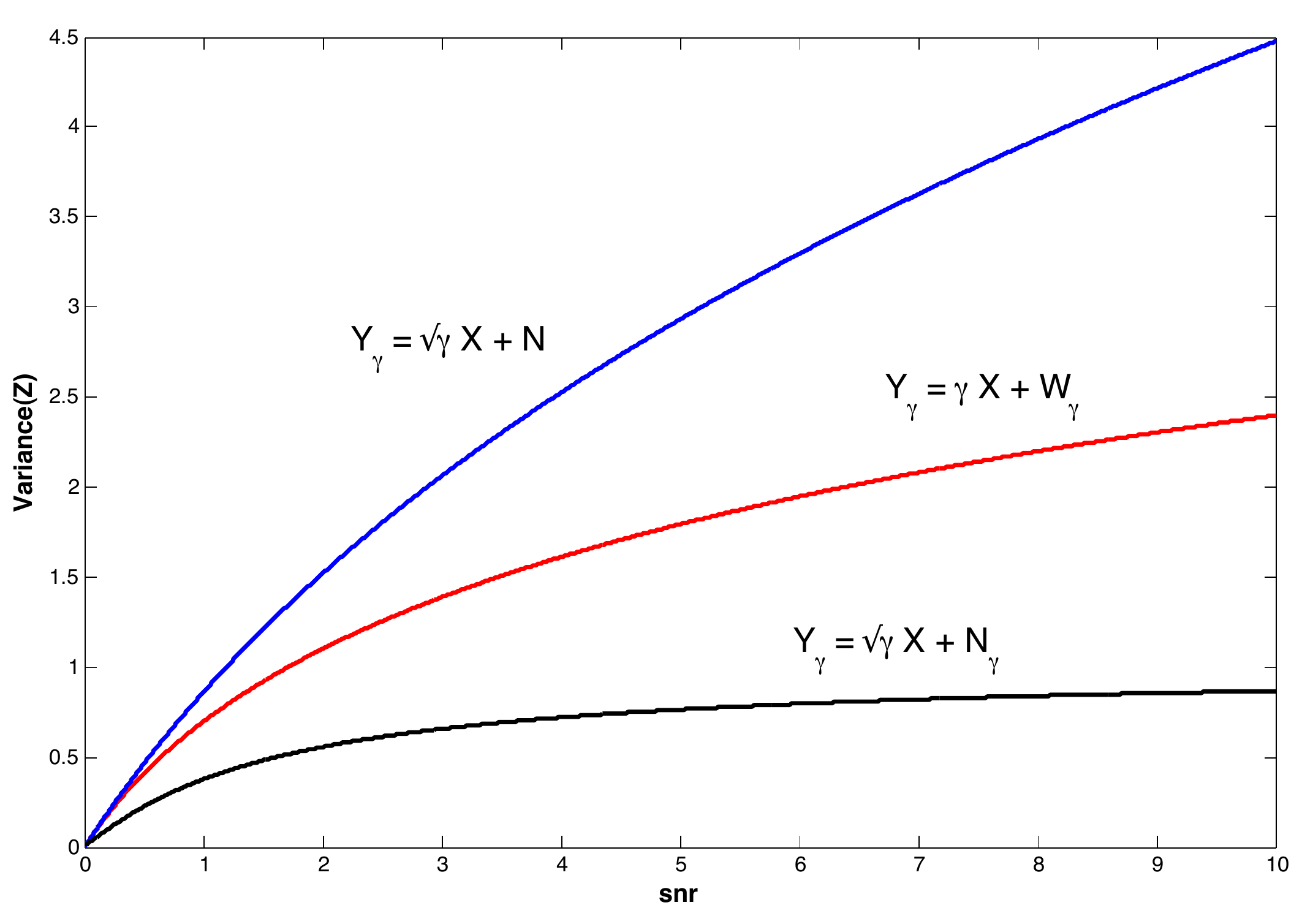}
\caption{Variance of Tracking Error $Z$ vs. snr for the cases in sections \ref{sec: scalar example 1}, \ref{sec: scalar example 2} and \ref{sec: scalar example 3}.}
\label{plot_var}
\end{center}
\end{figure}

Previously, in \ref{sec: scalar example 1}, \ref{sec: scalar example 2} and \ref{sec: scalar example 3} we have considered different couplings that are consistent with (\ref{eq: channel_cond}) and give rise to different pointwise relations between $X$ and $Y_0^\tsnr$. In particular, for the specific channel input $X \sim \Nscr(0,1)$, we have explicit characterizations of the tracking error $Z$ defined in Definition \ref{def: scalar setting definitions}. We have also calculated the variance of this tracking error for each of these process evolutions, and they are given by (\ref{eq: Z scalar variance for X Gaussian in Brownian motion coupling}), (\ref{eq: Z scalar variance for X Gaussian for Additive Standard Gaussian coupling}) and (\ref{eq: Z scalar variance for independent standard Gaussians coupling}) respectively. Here, we compare these couplings in terms of the variance of the tracking error for the Gaussian input. This comparison effectively tells us which particular relationship results in a better pointwise tracking of the information density and the actual squared error of the MMSE estimators. Fig.~\ref{plot_var} shows a plot of the error variances.
We observe that in this example of a Gaussian input, the coupling \ref{sec:
scalar example 3} results in the lowest variance of the tracking error, while
that of \ref{sec: scalar example 2} in the highest variance. We conjecture that
to be the case in general, i.e., for any distribution of $X$ with finite power.
\begin{remark}
Similar to our presentation of alternative couplings in the scalar estimation
problem, we could introduce a time-snr coupling of inputs and outputs in the
continuous time estimation model different from the Brownian sheet dependence in
Section \ref{sec: pointwise i-mmse for processes}. The difference in behavior
observed would be consistent with the differences we observe in the scalar
scenario. However, a detailed analysis of the continuous time I-MMSE
relationship for processes under alternative couplings is beyond the scope of
this paper. 
\end{remark}

\subsection{An identity}
As a final result, we present an interesting identity between two random quantities. The nature and applicability of such an identity needs to be explored further, but at the very least it shows us the kind of identities that can easily be reaped from our pointwise framework. Let $X \sim P$. We consider two different channels as follows:
\bea
Y_\gamma &=& \sqrt{\gamma} X + W_1 \label{eq: ch_1} \\
\tilde{Y}_\gamma &=& \gamma X + W_\gamma, \label{eq: ch_2} 
\eea
where $W_\gamma$ is a standard Brownian motion, independent of $X$. Thus, the coupling in (\ref{eq: ch_1}) is the ``Additive Standard Gaussian'' one of Subsection \ref{sec: scalar example 2} (with the role of $N$ played by $W_1$), while that in (\ref{eq: ch_2}) is our original Brownian motion coupling.  We note that
\bea \label{eq: pointwise equality of y}
Y_1 = \tilde{Y}_1
\eea
and consequently, 
\bea
\log \frac{d P_{Y_1|X}}{d P_{Y_1}} = \log \frac{d P_{\tilde{Y}_1|X}}{d P_{\tilde{Y}_1}} \ \ \ a.s. \label{eq: pointwise equality of RN derivative}
\eea
However, we are now in a position to invoke results derived in Section \ref{sec:
scalar matched setting} to characterize the quantities on either side of
(\ref{eq: pointwise equality of RN derivative}). Using the definition of $Z$ in
Definition \ref{def: scalar setting definitions} and Proposition \ref{lemma:
pointwise scalar Z for Brownian motion coupling}, we get respectively
\bea
Z_1 &=& \log \frac{d P_{Y_1|X}}{d P_{Y_1}} - \frac{1}{2} \int_0^1 (X - \E[X|Y_\gamma])^2 \,d\gamma \\
\tilde{Z}_1 &=& \log \frac{d P_{\tilde{Y}_1|X}}{d P_{\tilde{Y}_1}} - \frac{1}{2}
\int_0^1 (X - \E[X|\tilde{Y}_\gamma])^2 \,d\gamma = \int_0^1 (X -
\E[X|\tilde{Y}_\gamma]) \,dW_\gamma \ \ \ a.s.
\eea 
Combining the above we get  
\bea
Z_1 &=& \log \frac{d P_{Y_1|X}}{d P_{Y_1}} - \frac{1}{2} \int_0^1 (X - \E[X|Y_\gamma])^2 \,d\gamma \\
&=& \log \frac{d P_{\tilde{Y}_1|X}}{d P_{\tilde{Y}_1}} - \frac{1}{2} \int_0^1 (X - \E[X|Y_\gamma])^2 \,d\gamma \ \ \ a.s. \\
&=&  \int_0^1 (X - \E[X|\tilde{Y}_\gamma] \,dW_\gamma + \frac{1}{2} \int_0^1 (X - \E[X|\tilde{Y}_\gamma])^2 \,d\gamma - \frac{1}{2} \int_0^1 (X - \E[X|Y_\gamma])^2 \,d\gamma \ \ \ a.s. \label{eq: different channels pointwise}
\eea
where all equalities are valid in the almost sure sense.  Note that the relation in (\ref{eq: different channels pointwise}) is not only consistent with (and immediately implies) the I-MMSE relation, but also exhibits the pointwise relation between two different channel couplings. We can further plug in the expressions we have to characterize $Z_1$ (in subsection \ref{sec: scalar example 2}), to get equality relationships coupling the two channels considered.  

\section{Proofs} \label{sec: proofs}
Having discussed the  main results in Section \ref{sec: main results}, we now present the proofs in detail. We begin by proving our main result for the ``Pointwise Duncan'' setting in Proposition \ref{lemma: pointwise duncan using girsanov theory}. We then note that for the special case when the input is a DC process, the scalar estimation result in Proposition \ref{lemma: pointwise scalar Z for Brownian motion coupling} follows directly from the continuous-time result.
\begin{IEEEproof}[Proof of Proposition \ref{lemma: pointwise duncan using girsanov theory}]
We recall that the input-output mutual information is the expected value of the log Radon-Nikodym derivative of the measure induced by the process $\{Y^T\}$ conditioned on $X^T$ with respect to the measure induced by the process $\{Y^T\}$. The expectation is with respect to the law $P$. 

Define the causal estimate of $X_t$,
\bea
\hat{X}_t = \E[X_t | Y_0^t]. \label{eq: causal estimate of X}
\eea
We characterize the information density, by introducing the Radon-Nikodym
derivative with respect to the standard Wiener measure \cite{kailath71}. Using
the definition, and taking logarithm we get

\bea
\log \frac{d P_{Y^T|X^T}}{d P_{Y^T}} &=& \log \bigg({\frac{d P_{Y^T|X^T}}{d\mu}} \bigg{/} {\frac{dP_{Y^T}}{d\mu}} \bigg) \\
&=& \log {\frac{dP_{Y^T|X^T}}{d\mu}}  - \log{\frac{dP_{Y^T}}{d\mu}}. \label{eq: log_rn}
\eea

Let $\mu$ denote the standard Wiener measure on $Y^T$. We apply the Girsanov
theorem \cite{Girsanov60} to denote the Radon-Nikodym
derivatives of the conditional and marginal laws of $Y^T|X^T$ and $Y^T$
respectively with respect to $\mu$, as follows:
\bea
\log {\frac{dP_{Y^T|X^T}}{d\mu}} &=& \int_0^{T} X_t\,dY_t - \frac{1}{2} \int_0^{T}X_t^2 \,dt  \label{eq: conditional girsanov} \\
\log{\frac{dP_{Y^T}}{d\mu}} &=& \int_0^{T} \hat{X}_t \,dY_t - \frac{1}{2} \int_0^{T} (\hat{X}_t)^2\,dt, \label{eq: marginal girsanov}
\eea
where the equalities hold almost surely (a.s.). Therefore, from (\ref{eq: log_rn}), (\ref{eq: conditional girsanov}) and (\ref{eq: marginal girsanov}) we have
\bea
\log \frac{d P_{Y^T|X^T}}{d P_{Y^T}}  &=& \int_0^{T} \big(X_t - \hat{X}_t\big) \,dY_t - \frac{1}{2} \int_0^{T} \big(X_t^2 - (\hat{X}_t)^2 \big) \,dt. \label{eq: expression using girsanov}
\eea
We shall now proceed to simplify (\ref{eq: expression using girsanov}). Note from (\ref{eq: channel}) , (snr=1) 
\bea
\log \frac{d P_{Y^T|X^T}}{d P_{Y^T}} &=& \int_0^{T} \big(X_t - \hat{X}_t \big) \,(X_t\,dt + \,dW_t) - \frac{1}{2} \int_0^{T} \big(X_t^2 - (\hat{X}_t)^2 \big) \,dt \\
&=& \int_0^{T} \big(X_t^2 - X_t\hat{X}_t -\frac{1}{2}X_t^2 + \frac{1}{2}(\hat{X}_t)^2\big) \,dt + \int_0^{T} \big(X_t - \hat{X}_t\big)\,dW_t \\
&=& \frac{1}{2}\int_0^{T} \big(X_t - \hat{X}_t \big)^2 \,dt + \int_0^{T} \big(X_t - \hat{X}_t \big)\,dW_t.
\eea
On re-arranging, we get the desired result
\bea
D(T) &=& \int_0^{T} (X_t - \E[X_t|Y_0^t])\,dW_t  \ \ \ a.s.
\eea
\end{IEEEproof}
It is instructive to note that for the special case when $X_t \equiv X$ and $T=\tsnr$, Proposition \ref{lemma: pointwise duncan using girsanov theory} reduces directly to the scalar estimation setting in Section \ref{sec: scalar matched setting}, thereby giving a direct proof of Proposition \ref{lemma: pointwise scalar Z for Brownian motion coupling}. We now note that Theorem \ref{thm: variance D(T)} also follows directly from Proposition \ref{lemma: pointwise duncan using girsanov theory}, by using a familiar identity for It\^{o} integrals.
\begin{IEEEproof}[Proof of Theorem \ref{thm: variance D(T)}]  
\bea
Var (D(T)) &=& \E \big[ \{ D(T) \}^2\big] \label{eq: pf. variance D(T) (a)}\\
&=& \E \big[ \{\int_0^T (X_t - E[X_t | Y^t]) dW_t  \}^2\big] \\
&=& \int_0^T E [(X_t - E[X_t | Y^t])^2] dt. \label{eq: causal squared error at
snr=1} \\
&=& \text{cmmse}(T) = 2 I(X^T; Y^T)
\eea
Here, (\ref{eq: pf. variance D(T) (a)}) follows from the fact that $\E[D(T)] =
0$, and (\ref{eq: causal squared error at snr=1}) is a consequence of It\^{o}'s
Isometry property \cite[Chapter 6]{SteeleBook}. Note also, that by
definition, (\ref{eq: causal squared error at snr=1}) is the squared filtering
error, denoted by cmmse(T) in (\ref{eq: define cmmse(T)}), which in turn is
twice the mutual information by Duncan's theorem.
\end{IEEEproof}
We again observe that for the special choices $X_t \equiv X$ and $T=\tsnr$, we obtain the result for the scalar setting in Theorem \ref{thm: variance of Z scalar for Brownian motion coupling}, which tells us that the variance of the tracking error is equal to the minimum mean squared error integrated over SNR. Having established the pointwise results for the scalar and continuous time channels using Girsanov theory, we now proceed to prove the limit theorem that is a direct application of the pointwise treatment of Duncan's result.
\begin{IEEEproof}[Proof of Theorem \ref{thm: limit theorem}]
Note that Duncan's theorem tells us the equivalence of mutual information rate and half time-averaged causal squared error, i.e.
\bea
\frac{1}{T} I(X^T;Y^T) = \frac{1}{T} \int_0^T  \frac{1}{2} \E [ (X_t - E[X_t | Y_0^t])^2 ] \,dt.
\eea
Under the channel model in (\ref{eq: channel}), we can change the order of expectation to equivalently state the above identity as: 
\bea
\E \big[ \frac{1}{T} \log \frac{dP_{Y_0^T|X_0^T}}{dP_{Y_0^T}} \big] = \E \big[ \frac{1}{T} \int_0^T \frac{1}{2}(X_t - \E[X_t|Y_0^t])^2 \,dt \big]. \label{eq: normalized duncan}
\eea
Thus we observe the equality in expectation of the information density and half the squared filtering error. Using our pointwise characterization of Duncan's result, we will now show that not only are the random quantities in (\ref{eq: normalized duncan}) equal in expectation, but in the limit of large T, their difference converges to 0 in the mean square sense.
Let us recall (\ref{eq: duncan D(T)}) and the result in (\ref{eq: stochastic integral for Duncan}). On dividing both sides of (\ref{eq: stochastic integral for Duncan}) by $T$, we get the following:
\bea 
\frac{1}{T} \log \frac{dP_{Y_0^T|X_0^T}}{dP_{Y_0^T}} = \frac{1}{T} \int_0^T \frac{1}{2}(X_t - \E[X_t|Y_0^t])^2 \,dt +\frac{1}{T}  \int_0^T(X_t - E[X_t|Y_0^t])\cdot \,dW_t. \label{eq: time-averaged}
\eea
Let us take the limit as $T \rightarrow \infty$. We claim that under very basic regularity conditions on $X_t$, the second term in the r.h.s. goes to 0 in the mean square sense. 
\bea
\text{Var}\big(\frac{1}{T}  \int_0^T(X_t - E[X_t|Y_0^t])\cdot \,dW_t\big) &=& \frac{1}{T^2} \text{Var}\big(\int_0^T(X_t - E[X_t|Y_0^t])\cdot \,dW_t\big) \\
&=& \frac{1}{T^2} \int_0^T E [(X_t - E[X_t | Y^t])^2] dt. \label{eq: fractional variance of error}
\eea 
Thus, if the expression in (\ref{eq: fractional variance of error}) goes to 0 in the limit $T \rightarrow \infty$, i.e. it satisfies the condition in (\ref{eq: condition on growth rate}), then from (\ref{eq: time-averaged}) we get the desired result:
\bea
\text{l.i.m.}_{T \rightarrow \infty} \frac{1}{T} \left[  \log \frac{dP_{Y_0^T|X_0^T}}{dP_{Y_0^T}} - \frac{1}{2} \int_0^T (X_t - \E[X_t|Y_0^t])^2 \,dt \right]  = 0,  
\eea
where $\text{\text{\text{l.i.m.}}}$ denotes Limit in the Mean. 
\end{IEEEproof}
Having established the pointwise representation of Duncan's result and using it to extract results for scalar estimation, we now present the proof for continuous-time estimation with mismatch. We again note in this case, that the scalar estimation results follow as a special case of the continuous-time version of the result. 
\begin{IEEEproof}[Proof of Proposition \ref{lemma: mismatch M(T)} ]
Let us denote the causal estimates of $X_t$ under each law:
\bea
\pi^P_t = E_P[X_t | Y^t] \\
\pi^Q_t = E_Q[X_t | Y^t].
\eea
We first note that the innovations processes specified below,
\bea 
\bar{W}_t = Y_t - \int_0^t\pi^P_s \,ds
\eea 
and
\bea
\bar{V}_t = Y_t - \int_0^t\pi^Q_s \,ds
\eea 
are standard Brownian motions under $P$ and $Q$. We now apply Girsanov's theorem (also discussed in \cite[Section IV.D]{weissman10}) to characterize the log Radon-Nikodym derivative of the observed process under laws $P$ and $Q$ respectively,
\bea 
\log {\frac{dP_{Y^T}}{d\mu}} &=& \int_0^{T} \pi^P_t\,dY_t - \frac{1}{2} \int_0^{T}{(\pi^P_t)}^2 \,dt  \label{eq: girsanov on P} \\
\log{\frac{dQ_{Y^T}}{d\mu}} &=& \int_0^{T} \pi^Q_t \,dY_t - \frac{1}{2} \int_0^{T} {(\pi^Q_t)}^2\,dt. \label{eq: girsanov on Q}
\eea 
We now use (\ref{eq: girsanov on P}) and (\ref{eq: girsanov on Q}) to characterize the log Radon-Nikodym derivative of the measure induced by the output process $Y_0^T$ under $P$ with respect to the measure induced under $Q$, 
\begin{eqnarray}
\log \frac{d P_{Y_0^T}}{d Q_{Y_0^T}} &=& \log \bigg(\frac{d P_{Y_0^T}/d \mu}{d Q_{Y_0^T} / d \mu} \bigg) \\
&=& \int_0^T (\pi^P_t - \pi^Q_t) \,dY_t - \frac{1}{2} \int_0^T \big[ (\pi^P_t)^2 - (\pi^Q_t)^2\big] \,dt \\
&=& \int_0^T (\pi^P_t - \pi^Q_t) (X_t\,dt + \,dW_t) -
\frac{1}{2} \int_0^T \big[ (\pi^P_t)^2 - (\pi^Q_t)^2\big] \,dt \label{eq: pf.
RN (a)}\\
&=& \frac{1}{2}\int_0^T\big[- (\pi^P_t)^2 + 2 \pi^P_t X_t - 2 \pi^Q_t X_t+ (\pi^Q_t)^2\big]\,dt + \int_0^T(\pi^P_t - \pi^Q_t) \,dW_t \\
&=& \frac{1}{2}\int_0^T\bigg[(\pi^Q_t - X_t)^2 - (\pi^P_t - X_t)^2\bigg]\,dt + \int_0^T(\pi^P_t - \pi^Q_t) \,dW_t, \label{eq: getting expression for M(T)}
\end{eqnarray}
where (\ref{eq: pf. RN (a)}) follows from (\ref{eq: channel}).

Note from (\ref{eq: define M(T)}) and (\ref{eq: getting expression for M(T)}) that, $M(T)$ is a stochastic integral and is expressed as:
\begin{equation}
M(T) = \int_0^T(\pi^P_t - \pi^Q_t) \,dW_t \ \ \ P-a.s.
\end{equation} 
\end{IEEEproof}
We now invoke It\^{o}'s Isometry property along with Proposition \ref{lemma: mismatch M(T)} to prove the second moment result in Theorem \ref{thm: variance M(T)}.
\begin{IEEEproof}[Proof of Theorem \ref{thm: variance M(T)}]
 Using the isometry property of It\^{o} integrals, we can compute the variance of $M(T)$ as follows
\begin{eqnarray}
Var(M(T)) &=& E_P \bigg[ \int_0^T (\pi^P_t - \pi^Q_t)^2 \,dt \bigg] \\
&=& \int_0^T E_P \big[ (\pi^Q_t - X_t)^2 - (\pi^P_t-X_t)^2 \big]
\,dt \label{eq: pf. M(T) (a)}\\
&=& cmse_{P,Q}(T) - cmse_{P,P}(T), \\
&=& 2 D(P_{Y_0^T}||Q_{Y_0^T}) \label{eq: pf. M(T) (b)}
\end{eqnarray}
where (\ref{eq: pf. M(T) (a)}) follows from orthogonality property of
estimators, and (\ref{eq: pf. M(T) (b)}) follows from (\ref{eq: mismatched
result}).
\end{IEEEproof}
We now present the proofs of the pointwise results generalized to channels with feedback. The techniques are similar to the ones we use in order to prove the previous results for pointwise Duncan and mismatch. 
\begin{IEEEproof}[Proof of Theorem \ref{thm: pointwise result for phi channel}]
In proving the pointwise result for channels with feedback, we use the same idea that we employed in the pointwise treatment of Duncan's theorem. We use Girsanov theory to characterize the likelihood ratio of the conditional and marginal laws of $Y^T$, in terms of the filtering error in estimating $\phi_t$. Let $\mu$ be the standard Wiener measure. Then,
\bea
\log \frac{d P_{Y^T|X^T}}{d P_{Y^T}} &=& \log \bigg({\frac{d P_{Y^T|X^T}}{d\mu}} \bigg{/} {\frac{dP_{Y^T}}{d\mu}} \bigg) \\
&=& \log {\frac{dP_{Y^T|X^T}}{d\mu}}  - \log{\frac{dP_{Y^T}}{d\mu}} \ \ \ a.s. \label{eq: phi channel log RN}
\eea 
Using the formula for likelihood ratios in the presence of white noise (cf. \cite{kadota71},\cite{kailath71},\cite{kailath69}), we get the following almost-sure characterizations for the log Radon-Nikodym derivatives
\bea
\log {\frac{dP_{Y^T|X^T}}{d\mu}} &=& \int_0^{T} \phi_t\,dY_t - \frac{1}{2} \int_0^{T}\phi_t^2 \,dt \label{eq: conditional RN with phi}\\ 
\log{\frac{dP_{Y^T}}{d\mu}} &=& \int_0^{T} \hat{\phi}_t \,dY_t - \frac{1}{2} \int_0^{T} (\hat{\phi}_t)^2\,dt, \label{eq: marginal RN with phi}
\eea
where 
\bea
\hat{\phi}_t = \E[\phi_t | Y_0^t ], 
\eea
is the causal estimate of $\phi_t$ given the observations up until $t$.
From equations (\ref{eq: phi channel log RN}), (\ref{eq: conditional RN with phi}) and (\ref{eq: marginal RN with phi}), we get that 
\bea
\log \frac{d P_{Y^T|X^T}}{d P_{Y^T}} &=& \int_0^{T} \big(\phi_t - \hat{\phi}_t \big) \,(\,dY_t) - \frac{1}{2} \int_0^{T} \big(\phi_t^2 - (\hat{\phi}_t)^2 \big) \,dt \ \ \ a.s. \label{eq: information density for phi channel}
\eea
We recall that 
\bea
dY_t = \phi_t \,dt + \,dW_t,
\eea
and simplify  (\ref{eq: information density for phi channel}) to get 
\bea
\log \frac{d P_{Y^T|X^T}}{d P_{Y^T}} = \frac{1}{2}\int_0^{T} \big(\phi_t - \hat{\phi}_t \big)^2 \,dt + \int_0^{T} \big(\phi_t - \hat{\phi}_t \big)\,dW_t \ \ \ a.s. \label{eq: information density for phi channel simplified}
\eea
Recall that 
\bea
D_{\phi}(T) &\stackrel{\Delta}{=}& \log \frac{d P_{Y^T|X^T}}{d P_{Y^T}} - \frac{1}{2} \int_0^T (\phi_t - \hat{\phi}_t)^2 \,dt.
\eea
Combining with (\ref{eq: information density for phi channel simplified}), we get the desired result,
\bea
D_{\phi}(T) = \int_0^T(\phi_t - \hat{\phi}_t) \,dW_t \ \ \ a.s.
\eea
\end{IEEEproof}
We now use Theorem \ref{thm: pointwise result for phi channel} along with It\^{o}'s Isometry to directly yield the result in Corollary \ref{corollary: second moment phi channel}.
\begin{eqnarray}
Var(D_{\phi}(T)) &=& \E\big[ \big(\int_0^T(\phi_t - \hat{\phi}_t) \,dW_t\big)^2 \big] \\
&\stackrel{(a)}{=}&  \int_0^T \E[ (\phi_t - \hat{\phi}_t)^2 ] \,dt \\
&=& \mbox{$\sf{cmmse}$}_{\phi}(T),
\end{eqnarray}
where (a) follows from the Isometry property of stochastic integrals.

Note: Theorem \ref{thm: mismatch general channel} and Corollary \ref{corollary: second moment phi channel with mismatch} can be proved using the same techniques we used to establish the previous results for the non-mismatched case in Theorem \ref{thm: pointwise result for phi channel}, and Proposition \ref{lemma: mismatch M(T)} for the mismatched setting without feedback. Since the proofs follow directly, we omit them from the present discussion. 

We now explore the proof of the pointwise I-MMSE relationship for processes. We recall that in this setting, we have used a Brownian sheet process (\ref{eq: time-snr continuous channel}) to place the input and output processes on the same probability space on the time-snr plane. In this proof we use higher dimensional Girsanov theory to characterize the Radon-Nikodym derivative for vector processes. The proof is given below.  
\begin{IEEEproof}[Proof of Theorem \ref{thm: pointwise I-MMSE processes}]
 
We first present a result which can be established using Girsanov theory for higher dimensions for the continuous time Gaussian channel.

Let $\mathbf{X} \in \Re^M$ be a real-valued random vector governed by the law $P_{\mathbf{X}}$. This acts as input to a Gaussian channel (at SNR = 1) to yield the output $\mathbf{Y}_t$ as a function of observation time $t \in [0,T]$. The input and output are related by the following equation:
\bea
\mathbf{Y}_t = t\mathbf{X} + \mathbf{W}_t \hspace*{1em} t \in [0,\tsnr], \label{2_a_vec}
\eea
where $\mathbf{W}_t$ is a standard Brownian motion in $M$ dimensions. In this setting we are interested in the ``tracking error'' between the log Radon-Nikodym derivative and integral of half the squared error, in the time interval $[0,\tsnr]$. The corresponding definitions for the vector case are listed below:
\bea 
I_1 &=& \log \frac{dP_{\mathbf{Y}_{\tsnr}|\mathbf{X}}}{dP_{\mathbf{Y}_{\tsnr}}} \label{eq: vector defn pointwise1} \\ 
I_2 &=& \frac{1}{2} \int_0^\tsnr ||\mathbf{X} - \E[\mathbf{X}|\mathbf{Y}_0^t]||^2 \,dt \label{eq: vector defn pointwise2}
\eea  
\begin{lemma}
Let $\mathbf{Z} = I_1 - I_2$, where $I_1$ and $I_2$ are defined in (\ref{eq: vector defn pointwise1})-(\ref{eq: vector defn pointwise2}). Then,
\bea
\mathbf{Z} = \int_0^\tsnr ( \mathbf{X} - \E[\mathbf{X} | \mathbf{Y}_0^t)\cdot\,d\mathbf{W}_t, \ \ \ a.s. \label{pointwise identity in higher dimensions}
\eea
\end{lemma}
\begin{IEEEproof}
The proof is very similar to that of Theorem \ref{lemma: pointwise duncan using girsanov theory}. Let $\mu$ be the standard Wiener measure. We note that 
\bea
I_1 &=& \log \frac{dP_{\mathbf{Y}_{\tsnr}|\mathbf{X}}}{dP_{\mathbf{Y}_{\tsnr}}} \\
&=& \log \frac{dP_{\mathbf{Y}_{\tsnr}|\mathbf{X}}}{d\mu} - \log \frac{dP_{\mathbf{Y}_{\tsnr}}}{d\mu} \ \ \ a.s. \label{eq: I1 vector form}
\eea
Define
\bea
\hat{\mathbf{X}}_\gamma = \E[\mathbf{X} | \mathbf{Y}_0^\gamma ], 
\eea
for $\gamma \in [0,\tsnr]$. Now, using Girsanov theorem (cf. \cite[Section 3.5]{KaratzasShreve}) for higher dimensions, we can write the log Radon-Nikodym derivatives of the conditional and marginal distributions of $\mathbf{Y}$, with respect to $\mu$ as follows,
\bea
\log \frac{dP_{\mathbf{Y}_{\tsnr}|\mathbf{X}}}{d\mu} &=& \int_0^\tsnr \mathbf{X}\cdot \,d\mathbf{Y}_\gamma  - \frac{1}{2} \int_0^\tsnr \mathbf{X} \cdot \mathbf{X} \,d\gamma \ \ \ a.s. \\
\log \frac{dP_{\mathbf{Y}_{\tsnr}}}{d\mu} &=&  \int_0^\tsnr \hat{\mathbf{X}}_\gamma \cdot \,d\mathbf{Y}_\gamma  - \frac{1}{2} \int_0^\tsnr \hat{\mathbf{X}}_\gamma \cdot \hat{\mathbf{X}}_\gamma \,d\gamma \ \ \ a.s. 
\eea
Combining the above expressions with (\ref{eq: I1 vector form}) and using (\ref{2_a_vec}), we get
\bea
I_1 &=&  \int_0^\tsnr (\mathbf{X} - \hat{\mathbf{X}}_\gamma) \cdot \,d\mathbf{Y}_\gamma  - \frac{1}{2} \int_0^\tsnr (\mathbf{X} \cdot \mathbf{X} - \hat{\mathbf{X}}_\gamma \cdot \hat{\mathbf{X}}_\gamma)\,d\gamma,
\eea
which upon simplification reduces to,
\bea 
I_1 &=& \frac{1}{2} \int_0^\tsnr ||\mathbf{X} - \hat{\mathbf{X}}_\gamma||^2 \,d\gamma + \int_0^\tsnr (\mathbf{X} - \hat{\mathbf{X}}_\gamma) \cdot d \mathbf{W}_\gamma \ \ \ a.s. 
\eea
In other words,
\bea 
\mathbf{Z} &=& I_1 - I_2 \\
&=&  \int_0^\tsnr (\mathbf{X} - \hat{\mathbf{X}}_\gamma) \cdot d \mathbf{W}_\gamma \ \ \ a.s.
\eea
\end{IEEEproof}
Armed with this characterization in (\ref{pointwise identity in higher dimensions}), of the input-output information density in higher dimensions for the Gaussian channel in (\ref{eq: time-snr continuous channel}), we now proceed to establish a relationship akin to the I-MMSE for a specific class of processes. We begin by looking at piecewise constant scalar processes. The result we obtain (we shall argue), can be extended just as easily to the general class of finite power continuous-time processes using approximation arguments. 

Let $\tilde{X}_t$ be a continuous time process, and $\tilde{Y}_t$ be its noise corrupted version through the channel in (\ref{eq: channel}) observed in the interval $[0,T]$. Fix $M \in \N$. We let $\tilde{X}_t$ be a piecewise-constant process such that 
\bea
\tilde{X}_t \equiv X_i \hspace{1em} \text{for all } \frac{(i-1)}{M}T \leq t \leq \frac{i}{M}T,
\eea
where $X_i \sim P$, and $1 \leq i \leq M$. 

Fix $\tsnr > 0$. From \cite{gsv2005} we know that the mutual information is the integral over SNR of the smoothing error,
\bea
I(\tilde{X}_0^T;\tilde{Y}_0^{T,(\tsnr)}) = \frac{1}{2} \int_0^{\tsnr} (\tilde{X}_t - \E[\tilde{X}_t | \tilde{Y}_0^{T,(\gamma)} ])^2 \,d\gamma
\eea
Note that,
\bea
\E \bigg[ \log \frac{dP_{\tilde{Y}_0^{T,(\tsnr)} | \tilde{X}_0^T}}{dP_{\tilde{Y}_0^{T,(\tsnr)}}}\bigg] = I(\tilde{X}_0^T;\tilde{Y}_0^{T,(\tsnr)}) = I(\{X_i\}_{i=1}^{M};\{Y^{[i]}_{\tsnr}\}_{i=1}^{M})  = I(\mathbf{X};\mathbf{Y}_\tsnr) = \E \bigg[ \log \frac{dP_{\mathbf{Y}_{\tsnr}|\mathbf{X}}}{dP_{\mathbf{Y}_{\tsnr}}} \bigg], \label{mutual information of vector process}
\eea
where 
\begin{equation}
Y^{[i]}_\gamma = \gamma \frac{T}{M} X_i + W^{[i]}_\gamma, \label{eq: ith channel in pointwise i-mmse}
\end{equation}
and $\gamma \in [0,\tsnr]$ denotes the signal to noise ratio. 
\begin{itemize}
\item The random vectors $\mathbf{X} \in \R^{M}$ and $\mathbf{Y}_\gamma \in \R^{M}$ denote the collection of variables $\{X_i\}_{i=1}^{M}$ and $\{Y^{[i]}_\gamma\}_{i=1}^{M}$ respectively. 
\item $W^{[i]}_\gamma$ denotes the Brownian sheet process $W_{t}^{(\gamma)}$ (that drives the channel noise in (\ref{eq: time-snr continuous channel})), observed in the interval $[\frac{(i-1)}{M}T, \frac{i}{M}T]$. The M-dimensional random vector $\mathbf{W}_\gamma = \{W^{[i]}_\gamma \}_{i=1}^M$ denotes an M-dimensional Brownian motion indexed by $\gamma$. 
\end{itemize}
Note that $\log \frac{dP_{\tilde{Y}_0^T | \tilde{X}_0^T ; \tsnr}}{dP_{\tilde{Y}_0^T ; \tsnr}}$ depends only on the joint distribution between $\{X_i\}_{i=1}^{M}$ and $\{Y^{[i]}_{\tsnr}\}_{i=1}^{M}$. Therefore it is the same as $\log \frac{dP_{\mathbf{Y}_{\tsnr}|\mathbf{X}}}{dP_{\mathbf{Y}_{\tsnr}}}$, not only under expectation, but also pointwise a.s., i.e.
\bea
 \log \frac{dP_{\tilde{Y}_0^{T,(\tsnr)} | \tilde{X}_0^T}}{dP_{\tilde{Y}_0^{T,(\tsnr)}}} = \log \frac{dP_{\mathbf{Y}_{\tsnr}|\mathbf{X}}}{dP_{\mathbf{Y}_{\tsnr}}} \ \ \ a.s. \label{eq: log rn of vector}
\eea

We now apply the identity in (\ref{pointwise identity in higher dimensions}) directly to the log Radon-Nikodym derivative (quantity inside brackets in r.h.s. of (\ref{eq: log rn of vector})) to get
\bea
\log \frac{dP_{\mathbf{Y}_{\tsnr}|\mathbf{X}}}{dP_{\mathbf{Y}_{\tsnr}}} &=& \frac{1}{2} \int_0^\tsnr ||\mathbf{X} - \E[\mathbf{X}|\mathbf{Y} ; \gamma]||^2 \,d\gamma + \int_0^\tsnr ( \mathbf{X} - \E[\mathbf{X} | \mathbf{Y}_\gamma] )\cdot\,d\mathbf{W}_\gamma \\
&=& \frac{1}{2} \int_0^\tsnr \int_0^T (\tilde{X}_t - \E[\tilde{X}_t | \tilde{Y}_0^T ; \gamma])^2 \,dt \,d\gamma + \int_0^\tsnr ( \mathbf{X} - \E[\mathbf{X} | \mathbf{Y}_\gamma] )\cdot\,d\mathbf{W}_\gamma \label{eq: intermediate step in pointwise I-MMSE}\\
&=& \frac{1}{2} \int_0^\tsnr \text{sse(T,}\gamma) \,d\gamma + \int_0^\tsnr \sum_{i=1}^{M} (X_i - \E[X_i | \mathbf{Y} ; \gamma])\,dW^{[i]}_\gamma 
\eea
where sse(T,$\gamma$) represents the {\em squared smoothing error} in an observation window of duration T at signal to noise ratio $\gamma$. Note that $\E[\text{sse(T,}\gamma)] = \text{mmse}(\gamma)$.

We can write the difference as, 
\bea
 N(T) \stackrel{\Delta}{=} \log \frac{dP_{\tilde{Y}_0^{T,(\tsnr)} | \tilde{X}_0^T}}{dP_{\tilde{Y}_0^{T,(\tsnr)}}} - \frac{1}{2} \int_0^\tsnr \text{sse(T,}\gamma) \,d\gamma  &=& \int_0^\tsnr \sum_{i=1}^{M} (X_i - \E[X_i | \mathbf{Y} ; \gamma])\,dW^{[i]}_\gamma \\  &=&  \sum_{i=1}^{M} \int_0^\tsnr (X_i - \E[X_i | \tilde{Y}_0^T ; \gamma])\,dW^{[i]}_\gamma 
\eea
Note that by It\^{o}'s Isometry property,
\bea
Var(N(T)) &=& \int_0^{\tsnr} \E\big[ ||\mathbf{X} - \E[\mathbf{X}|\mathbf{Y} ; \gamma]||^2 \big]\,d\gamma \\
&=& \int_0^\tsnr \E \bigg[\int_0^T (\tilde{X}_t - \E[\tilde{X}_t | \tilde{Y}_0^T ; \gamma])^2 \,dt\bigg] \,d\gamma \\
&=& \int_0^{\tsnr} \text{mmse}(\gamma) \stackrel{(f)}{=} 2 I(\tsnr),
\eea
which is also an interesting result. Note that (f) follows from the
continuous-time I-MMSE relationship. We now argue that the result that we just
established for piecewise constant processes carries through for general
continuous time processes with finite average power. We refer the reader to
\cite[Section IV.C]{weissman10} for details in making this approximation, and
provide the sketch below.

The main idea is to induce a stepwise process $X_0^{(n),T}$ defined by
\bea 
X_t^{(n)} \equiv \frac{1}{2^n T}\int_{i2^{-n}T}^{(i+1)2^{-n}T} X_t \,dt \text{ for } t \in (i2^{-n}T, (i+1)2^{-n}T ].
\eea
Since processes in $P$ have finite energy, the integral $\int_{i2^{-n}T}^{(i+1)2^{-n}T} X_t \,dt$ exists and is finite P-a.s. and is in $L_2(P)$. We now note that,
\bea
X_0^{(n),T} \stackrel{n \to \infty}{\rightarrow} X_0^T \text{ in } L^2(\,dt \,dP).
\eea
Further, the Radon-Nikodym derivates of the induced measures in the stepwise process converge to the actual Radon-Nikodym derivates in a $P_{{Y}_0^T}$-a.s. sense. Therefore, the approximation allows us to generalize our results to the class of all finite power continuous time processes. 
\end{IEEEproof}
Duncan's theorem proves as a corollary, the equivalence of the causal and anti-causal squared errors. We now present a proof for Proposition \ref{lemma: difference of causal anti-causal errors}, where we establish a pointwise version of the result.
\begin{IEEEproof}[Proof of Proposition \ref{lemma: difference of causal anti-causal errors}]
We first note the following relationship which follows directly from Theorem \ref{lemma: pointwise duncan using girsanov theory},
\bea
\log \frac{dP_{Y_0^T|X_0^T}}{dP_{Y_0^T}} = \frac{1}{2}\int_0^T (X_t - \E[X_t|Y_0^t])^2 \,dt + \int_0^T(X_t - E[X_t|Y_0^t])\cdot \,dW_t \ \ \ a.s. \label{eq: pointwise logRN and cmse}
\eea 
Let us recall the transformations defined in (\ref{eq: tilded process for causal anticausal error beg}) - (\ref{eq: tilded process for causal anticausal end}).
Note in addition, that  $\E[ \tilde{X}_t| \tilde{Y}_0^t]$ is adapted to the filtration induced by $(\tilde{X}^T, B^T)$, so
\bea \label{eq: legit stoc integral}
 \int_0^T  (\tilde{X}_t - E[ \tilde{X}_t| \tilde{Y}_0^t]) \cdot dB_t
\eea
is well defined in the standard sense of an It\^{o} integral. 

Applying the relation in (\ref{eq: pointwise logRN and cmse}), with the associations $X^T \rightarrow \tilde{X}^T$, $W^T \rightarrow B^T$ and $Y^T \rightarrow \tilde{Y}^T$, gives us
\bea \label{eq: log likelihood result time reversed}
\log \frac{dP_{\tilde{Y}_0^T| \tilde{X}_0^T}}{dP_{\tilde{Y}_0^T}} = \frac{1}{2}\int_0^T ( \tilde{X}_t - \E[ \tilde{X}_t| \tilde{Y}_0^t])^2 \,dt + \int_0^T( \tilde{X}_t - E[ \tilde{X}_t| \tilde{Y}_0^t])\cdot \,dB_t  \ \ \ a.s. 
\eea
On the other hand, since there are one-to-one transformations to get from  $\tilde{X}^T$ to $X^T$ and from $\tilde{Y}^T$ to $Y^T$, we have 
\bea \label{eq: log likelihood result original}
\log \frac{dP_{Y_0^T|X_0^T}}{dP_{Y_0^T}} = \log \frac{dP_{\tilde{Y}_0^T| \tilde{X}_0^T}}{dP_{\tilde{Y}_0^T}} \ \ \ a.s.
\eea
Combining (\ref{eq: log likelihood result original}), (\ref{eq: log likelihood result time reversed}) and (\ref{eq: pointwise logRN and cmse}), we get the following equality (in the almost sure sense):
\begin{align} \label{eq: difference of errors} 
 \frac{1}{2}\int_0^T (X_t - \E[X_t|Y_0^t])^2 \,dt -  \frac{1}{2} \int_0^T (
\tilde{X}_t - \E[ \tilde{X}_t| \tilde{Y}_0^t])^2 \,dt  =  \int_0^T
(\tilde{X}_t -  E[ \tilde{X}_t| \tilde{Y}_0^t] ) \cdot dB_t  - \int_0^T (X_t -
E[X_t|Y_0^t] ) \cdot dW_t  
\end{align}

Simplifying the above expression further, we get
\begin{align} \label{eq: pointwise between causal and noncausal}
 \frac{1}{2}\int_0^T (X_t - \E[X_t|Y_0^t])^2 \,dt -  \frac{1}{2} \int_0^T ( \tilde{X}_t - \E[ \tilde{X}_t| \tilde{Y}_0^t])^2 \,dt   &=  \int_0^T( \tilde{X}_t - E[ \tilde{X}_t| \tilde{Y}_0^t])\cdot \,dB_t - \int_0^T(X_t - E[X_t|Y_0^t])\cdot \,dW_t  \\  
 &= \int_0^T  E[X_t|Y_0^t]  dW_t -  \int_0^T  E[ \tilde{X}_t| \tilde{Y}_0^t]) dB_t ,
\end{align}
where the second equality holds true for all processes that satisfy the following benign condition in (\ref{eq: eq stoch integral}): 
\bea \label{eq: eq stoch integral}
\int_0^T   \tilde{X}_t  dB_t = \int_0^T X_t  dW_t     \ \ \ a.s.
\eea
Note that (\ref{eq: eq stoch integral}) is simple to verify for the class of piecewise constant processes on the interval $[0,T]$. By approximation arguments similar to the ones given in the proof of Theorem \ref{thm: pointwise I-MMSE processes}, one can establish the equality (\ref{eq: eq stoch integral}) for all finite power continuous time processes. 
\end{IEEEproof}

Having established the pointwise I-MMSE relationship for continuous-time processes, we now present a pointwise version of the causal vs. non-causal error relationship discussed in Subsection \ref{sec: pointwise causal vs. non-causal error}. In doing so, for simplicity, we restrict our attention to the class of piecewise-constant processes. Our characterization is valid for all piecewise constant processes, and (appealing to the arguments given in the proof of Theorem \ref{thm: pointwise I-MMSE processes}) therefore holds for all continuous time processes that can be approxmiated as such.

Let $\tilde{X}_0^T$ denote a piecewise-constant process, such that 
\bea
\tilde{X}_t \equiv X_i \hspace{1em} \text{for all } \frac{(i-1)}{M}T \leq t \leq \frac{i}{M}T,
\eea
where $X_i \sim P$, and $1 \leq i \leq M$. We fix snr=1, and let the output be denoted by $\{\tilde{Y}_t^{(\gamma)}\}_{0 \leq t \leq T}$ for the channel described in (\ref{eq: time-snr continuous channel}). I.e.,
\bea
d\tilde{Y}^{(\gamma)}_t = \gamma \tilde{X}_t\,dt + dW^{(\gamma)}_t,
\eea
for $\gamma \in [0,1]$ and $t \in [0,T]$. We first note from Proposition \ref{lemma: pointwise duncan using girsanov theory} that the input-output information density can be written as
\bea
\log \frac{dP_{\tilde{Y}_0^{T,(1)}|\tilde{X}_0^T}}{dP_{\tilde{Y}_0^{T,(1)}}} = \frac{1}{2}\int_0^T (\tilde{X}_t - \E[\tilde{X}_t|\tilde{Y}_0^{t,(1)}])^2 \,dt + \int_0^T(\tilde{X}_t - E[\tilde{X}_t|\tilde{Y}_0^{t,(1)}])\cdot \,dW^{(1)}_t \ \ \ a.s. \label{eq: information density via pointwise duncan}
\eea 
Recall the definitions of $\mathbf{X}$, $\mathbf{Y}_\gamma$ and $\mathbf{W}_\gamma$ in the discussion following (\ref{eq: ith channel in pointwise i-mmse}) in the proof of Theorem \ref{thm: pointwise I-MMSE processes}. Also, in establishing Theorem \ref{thm: pointwise I-MMSE processes}, we derive a relationship for the input-output information density from (\ref{eq: log rn of vector}) and (\ref{eq: intermediate step in pointwise I-MMSE}), namely:
\bea 
\log \frac{dP_{\tilde{Y}_0^{T,(1)} | \tilde{X}_0^T}}{dP_{\tilde{Y}_0^{T,(1)}}} = \frac{1}{2} \int_0^1 \int_0^T (\tilde{X}_t - \E[\tilde{X}_t | \tilde{Y}_0^{T,(1)}])^2 \,dt \,d\gamma + \int_0^1 ( \mathbf{X} - \E[\mathbf{X} | \mathbf{Y}_\gamma] )\cdot\,d\mathbf{W}_\gamma \ \ \ a.s. \label{eq: information density via pointwise i-mmse}
\eea
We now combine (\ref{eq: information density via pointwise duncan}) and (\ref{eq: information density via pointwise i-mmse}) to write down the pointwise characterization of the filtering and smoothing errors,
\bea
\int_0^T (\tilde{X}_t - \E[\tilde{X}_t|\tilde{Y}_0^{t,(1)}])^2 \,dt - \int_0^1 \int_0^T (\tilde{X}_t - \E[\tilde{X}_t | \tilde{Y}_0^{T,(\gamma)}])^2 \,dt \,d\gamma = \nonumber \\
\int_0^1 2( \mathbf{X} - \E[\mathbf{X} | \mathbf{Y}_\gamma] )\cdot\,d\mathbf{W}_\gamma - \int_0^T2(\tilde{X}_t - E[\tilde{X}_t|\tilde{Y}_0^{t,(1)}])\cdot \,dW^{(1)}_t \ \ \ a.s. \label{eq: pointwise causal non-causal}
\eea 
Thus, the difference between the filtering and smoothing errors has an explicit characterization in terms of stochastic integrals. It is instructive to note that taking expectation on both sides of (\ref{eq: pointwise causal non-causal}) establishes the identity in (\ref{eq: causal vs. non-causal relationship}), for snr=1, by using the fact that the It\^{o} integrals in the r.h.s. of (\ref{eq: pointwise causal non-causal}) have zero mean.  

We now present proofs of results stated in Section \ref{sec: revisiting the scalar setting}. In Proposition \ref{lemma: conditional expectation of Z scalar}, we established that $\E[Z|X] = 0$, for all underlying distributions of the signal $X$ that have finite variance. The proof invokes the setting in \cite{ver_mis10} and is presented below. 
\begin{IEEEproof}[Proof of Proposition \ref{lemma: conditional expectation of Z scalar}]
We denote the mean squared error due to mismatch at signal to noise ratio $\gamma > 0$, by $\text{mse}_{P,Q}(\gamma)$ defined in (\ref{eq: mismatched mse}). Note from \cite[Section V.]{ver_mis10} that we have, 
\bea
I(X; \sqrt{\tsnr}X + N) \\
&=& \int D(N + \sqrt{\tsnr}x || N + \sqrt{\tsnr}X)\,dP_X(x) \label{eq: relative entropy form of I}\\
&=& \int D(\delta_x \ast \Nscr(0,\tsnr^{-1}) || P_X \ast \Nscr(0,\tsnr^{-1}))\, dP_X(x) \nonumber \\
&=& \frac{1}{2} \int \int_0^{\tsnr} \text{mse}_{P_x,\delta_x}(\gamma) - \text{mmse}_{\delta_x}(\gamma)\,d\gamma \,dP_X \\ 
&=& \frac{1}{2} \int \int_0^{\tsnr} \text{mse}_{P_x,\delta_x}(\gamma)\,d\gamma\,dP_X \label{eqn:ver1} \\
&=& \frac{1}{2}\int_0^{\tsnr} \text{mmse}(\gamma)\,d\gamma
\eea
From (\ref{eq: relative entropy form of I}) and (\ref{eqn:ver1}) it is clear that 
\bea
\E\bigg[\log\frac{dP_{Y_{\tsnr}|X}}{dP_{Y_{\tsnr}}} \big{|} X\bigg] &=& \E\bigg[ \frac{1}{2}\int_0^{\tsnr}(X - \E[X|Y_{\gamma}])^{2}\,d\gamma \big{|} X\bigg] \\
\implies E[Z|X] &=& 0 \ \ \ a.s.\label{cond_id}
\eea\bea
I_1 &=& \frac{1}{2}\log(1+\tsnr) + \frac{1}{2}\bigg(\frac{Y^2}{1 + \tsnr} - (Y - \sqrt{\tsnr}X)^2 \bigg) \\
&=& \frac{1}{2}\log(1+\tsnr) + \frac{1}{2}\bigg(\frac{\tsnr}{1+\tsnr}(X^2 - N^2) + 2XN\frac{\sqrt{\tsnr}}{1+\tsnr} \bigg) \label{z3_g1}
\eea
Also,
\bea
\E[I_1|X] = \frac{1}{2}\log(1+\tsnr) + \frac{1}{2}\bigg(\frac{\tsnr}{1+\tsnr}(X^2 - 1) \bigg) \label{z3_g2}
\eea
for all underlying processes $P_X$. Thus, not only are the random quantities above equal in expectation, but they are also equal in conditional expectation on $X$. 
\end{IEEEproof}

We now turn our attention to the alternative coupling discussed in Section \ref{sec: scalar example 2}. In Lemma \ref{lemma: Z scalar additive standard Gaussian}, we derived an expression for the pointwise tracking error for the Additive Standard Gaussian coupling (\ref{eq: channel in example 1})for a general input signal $X$. In the following, we present a proof of Lemma \ref{lemma: Z scalar additive standard Gaussian}.
\begin{IEEEproof}[Proof of Lemma \ref{lemma: Z scalar additive standard Gaussian}]
 
Let us assume $X$ is distributed according to $P_X$. Let $f_Z(z)$ denote the
probability density function of a standard Gaussian random variable. Note for
the setting in (\ref{eq: channel in example 1}) that,
\bea
P_{Y_\gamma|X}(y|x) &=& \frac{1}{\sqrt{2 \pi}} e^{-\frac{1}{2}(y-\sqrt{\gamma}x)^2} = f_Z(y - \sqrt{\gamma}x),
\eea
and
\bea 
P_{Y_\gamma}(y) &=& \int P_{Y_\gamma|X}(y|\tilde{x}) \,d P_{\tilde{x}}. 
\eea
Then,
\bea
\label{eq: I1 density form}
I_1(X,Y_\gamma) &=& \log \frac{P_{Y_\gamma|X}(Y_\gamma|X)}{P_{Y_\gamma}(Y_\gamma)}. 
\eea
We now look at the differential form of the I-MMSE relationship. Namely,
\bea
\frac{\partial I(X;Y_{\gamma})}{\partial \gamma} &=& \frac{1}{2} \text{mmse}(\gamma), 
\eea
or its pointwise equivalent
\bea \label{eq: pointwise differential I-MMSE}
\E \big[ \frac{\partial}{\partial \gamma} \log \frac{P_{Y_\gamma|X}(Y_\gamma|X)}{P_{Y_\gamma}(Y_\gamma)} - \frac{1}{2} (X-\E[X|Y_{\gamma}])^2 \big] = 0.
\eea
Define 
\bea
\tilde{Z}_\gamma =   \frac{\partial}{\partial \gamma} \log \frac{P_{Y_\gamma|X}(Y_\gamma|X)}{P_{Y_\gamma}(Y_\gamma)} - \frac{1}{2} (X-\E[X|Y_{\gamma}])^2. \label{eq: define Z tilde gamma}
\eea 
Differentiating (\ref{eq: I1 density form}) w.r.t. snr, we have
\bea
\frac{\partial}{\partial \gamma} \log \frac{P_{Y_\gamma|X}(Y_\gamma|X)}{P_{Y_\gamma}(Y_\gamma)} &=& \frac{\partial}{\partial \gamma} \log P_{Y_\gamma|X}(Y_\gamma|X) - \frac{\partial}{\partial \gamma} \log P_{Y_\gamma}(Y_\gamma) \\
 &=& 0 - \frac{\partial}{\partial \gamma} \log
P_{Y_\gamma}(Y_\gamma) \label{eq: pf. standard gaussian (a)}\\
 &=& \frac{1}{P_{Y_\gamma}(Y_\gamma)} \int \frac{1}{\sqrt{2
\pi}}e^{-\frac{1}{2} (Y_\gamma - \sqrt{\gamma}\tilde{x})^2}(Y_\gamma -
\sqrt{\gamma}\tilde{x}) ((X - \tilde{x})/{2\sqrt{\gamma}}) \,d P_{\tilde{x}}
\label{eq: pf. standard gaussian (b)}\\
 &=& \frac{1}{2\sqrt{\gamma}}\int \frac{P_{Y_\gamma|X}(Y_\gamma | \tilde{x})}{P_{Y_\gamma}(Y_\gamma)} (X Y_\gamma - Y_\gamma \tilde{x} - \sqrt{\gamma}X \tilde{x} + \sqrt{\gamma} \tilde{x}^2) \,d P_{\tilde{x}} \\
 &=& \frac{1}{2\sqrt{\gamma}} \big[XY_\gamma - (\sqrt{\gamma}X +
Y_\gamma)\E[X|Y_\gamma] + \sqrt{\gamma}\E[X^2|Y_\gamma] \big], \label{eq: I1
derivative w.r.t. snr}
\eea
where:
\begin{itemize}
 \item (\ref{eq: pf. standard gaussian (a)}) $\Leftarrow$
$P_{Y_\gamma|X}(Y_\gamma|X) = f_Z(Y_\gamma - \sqrt{\gamma}X) = f_Z(N)$ (and
hence does not depend on $\gamma$)
 \item (\ref{eq: pf. standard gaussian (b)}) $\Leftarrow$ $Y_\gamma =
\sqrt{\gamma}X + N$ is the explicit dependence of $Y_\gamma$ on $\gamma$. Using
this, and differentiating with respect to $\gamma$, we obtain the required
expression
 \item (\ref{eq: I1 derivative w.r.t. snr}) $\Leftarrow$ the integral is with
respect to the conditional law $X|Y_\gamma$, with $\tilde{x}$ as the variable of
integration, while keeping $X$ and $Y_\gamma$ constant.
\end{itemize}

Combining (\ref{eq: pointwise differential I-MMSE}), (\ref{eq: I1 derivative w.r.t. snr}) and (\ref{eq: define Z tilde gamma}), we get  
\bea
\tilde{Z}_\gamma &=& \frac{\partial I_1}{\partial \gamma} - \frac{1}{2} (X-\E[X|Y_{\gamma}])^2 \\
&=& \frac{1}{2\sqrt{\gamma}} \big[XY_\gamma - (\sqrt{\gamma}X + Y_\gamma)\E[X|Y_\gamma] + \sqrt{\gamma}\E[X^2|Y_\gamma] \big] - \frac{1}{2} (X-\E[X|Y_{\gamma}])^2 \\
&=& \frac{1}{2}\big{\{}\E[X^2|Y_\gamma] - X\E[X|Y_\gamma] - (X - \E[X|Y_\gamma])^2 + \frac{Y_\gamma}{\sqrt{\gamma}}(X - \E[X|Y_\gamma])\big{\}}.
\eea
Note now that $Z$ according to Definition \ref{def: scalar setting definitions} is
\bea
Z &=& \log \frac{P_{Y_\tsnr|X}(Y_\tsnr|X)}{P_{Y_\tsnr}(Y_\tsnr)} - \frac{1}{2} \int_0^\tsnr (X - \E[X|Y_\gamma)^2 \,d\gamma \\
&=& \int_0^{\tsnr} \tilde{Z}_\gamma \,d\gamma. 
\eea
\end{IEEEproof}

\section{Conclusion} \label{sec: conclusion}
We consider the scenario of mean square estimation of a signal observed through additive white Gaussian noise. We formulate classical information and estimation relationships in these contexts as expectation identities. We explicitly characterize the input-output information density for both scalar and continuous time Gaussian channels. Using this characterization, which relies on Girsanov theory, we obtain pointwise representations of these identities with the expectations removed and discover that these random quantities also have classical information-estimation links. In particular, canonical measures of information and estimation appear to be bridged by the second moment of the pointwise tracking error between the information density and the scaled filtering error. In this manner we present pointwise relations for Duncan's theorem, mismatched estimation, channels with feedback, the I-MMSE relationship as well as the causal vs. non-causal and causal vs. anticausal errors. A special treatment for scalar estimation is also provided where we present and discuss alternative couplings to the Brownian motion corrupted channel. We also provide applications of these results to obtain new and interesting relations in the information-estimation arena. 

The first and second moments of the tracking error in the Gaussian setting have direct
implications on information and estimation relations. In future work,we would
like to see whether similar implications emerge for higher order moments as
well. In addition it would be interesting to investigate whether pointwise relationships similar to the Gaussian case, hold also for the Poissonian
channel,
where links between estimation and information have been recently
uncovered in \cite{WeissmanAtar11} for a natural loss function. 

\section*{Acknowledgement}
The authors thank Rami Atar for valuable discussions. This work has been supported under a Stanford Graduate Fellowship, NSF grant CCF-0729195, and the Center for Science of Information (CSoI), an NSF Science and Technology Center, under grant agreement CCF-0939370.
\bibliographystyle{IEEEtran}
%

\end{document}